\documentclass[utf8]{nobranding} 

\usepackage{url,hyperref,lineno,microtype,subcaption}
\hypersetup{hidelinks}
\usepackage{wrapfig}
\usepackage{siunitx}
\usepackage{setspace}
\usepackage{tikz}
\usepackage{onimage}

\def\keyFont{\fontsize{8}{11}\helveticabold }
\def\firstAuthorLast{R. M. Dorrell {et~al.}}
\def\Authors{
    Robert M. Dorrell$^{1,*}$,
    Charlie J. Lloyd$^1$, 
    Ben J. Lincoln$^2$,
    Tom P. Rippeth$^2$, 
    John R. Taylor$^3$, 
    Colm-cille P. Caulfield$^{3,4}$, 
    Jonathan Sharples$^5$, 
    Jeff A. Polton$^6$,
    Brian D. Scannell$^2$,
    Deborah M. Greaves$^7$, 
    Rob A. Hall$^8$,
    John H. Simpson$^2$
}
\def\Address{
    $^1$ Energy and Environment Institute, University of Hull, HU6 7RX, UK.\\%
    $^2$ School of Ocean Sciences, Bangor University, LL59 5AB, UK.\\
    $^3$ Department of Applied Mathematics and Theoretical Physics, University of Cambridge, CB3 0WA, UK.\\%
    $^4$ BP Institute, University of Cambridge, Madingley Rise, Madingley Road, Cambridge, CB3 0EZ, UK.\\%
    $^5$ School of Environmental Sciences, University of Liverpool, Liverpool, L3 5DA, UK.\\%
    $^6$ National Oceanography Center, Joseph Proudman Building, Liverpool, L3 5DA, UK.\\%
    $^7$ School of Engineering, Computing and Mathematics, University of Plymouth, Drake Circus, Plymouth, PL4 8AA, UK.\\%
    $^8$ Centre for Ocean and Atmospheric Sciences, School of Environmental Sciences, University of East Anglia, NR4 7TJ, UK.
}

\def\corrAuthor{Robert M. Dorrell}
\def\corrEmail{r.dorrell@hull.ac.uk}

\begin{document}
\onecolumn
\firstpage{1}
\title[Infrastructure Mixing of Shelf Seas]{Anthropogenic Mixing of Seasonally Stratified Shelf Seas by Offshore Wind Farm Infrastructure} 
\author[\firstAuthorLast ]{\Authors} 
\address{} 
\correspondance{} 
\extraAuth{}%
\maketitle
\begin{abstract}
The offshore wind energy sector has rapidly expanded over the past two decades, providing a renewable energy solution for coastal nations. 
Sector development has been led in Europe, but is growing globally. 
Most developments to date have been in well-mixed, i.e. unstratified, shallow-waters near to shore. 
Sector growth is, for the first time, pushing developments to deep water, into a brand new environment: seasonally stratified shelf seas. 
Seasonally stratified shelf seas, where water density varies with depth, have a disproportionately key role in primary production, marine ecosystem and biochemically cycles. 
Infrastructure will directly mix stratified shelf seas.
The magnitude of this mixing, additional to natural background processes, has yet to be fully quantified. If large enough it may erode shelf sea stratification.
Therefore, offshore wind growth may destabilize and fundamentally change shelf sea systems.
However, enhanced mixing may also positively impact some marine ecosystems. 
This paper sets the scene for sector development into this new environment, reviews the potential physical and environmental benefits and impacts of large scale industrialization of seasonally stratified shelf seas and identifies areas where research is required to best utilise, manage and mitigate environmental change.
\newline
\tiny
 \keyFont{ \section{Keywords:} offshore wind energy, shelf seas, marine biogeochemistry, stratification, turbulent mixing} 
\end{abstract}
\section{Introduction}
\begin{figure}
\begin{tikzpicture}
\centering
\footnotesize
\node[draw=none,fill=none] at (0,0){\includegraphics[width=0.975\textwidth]{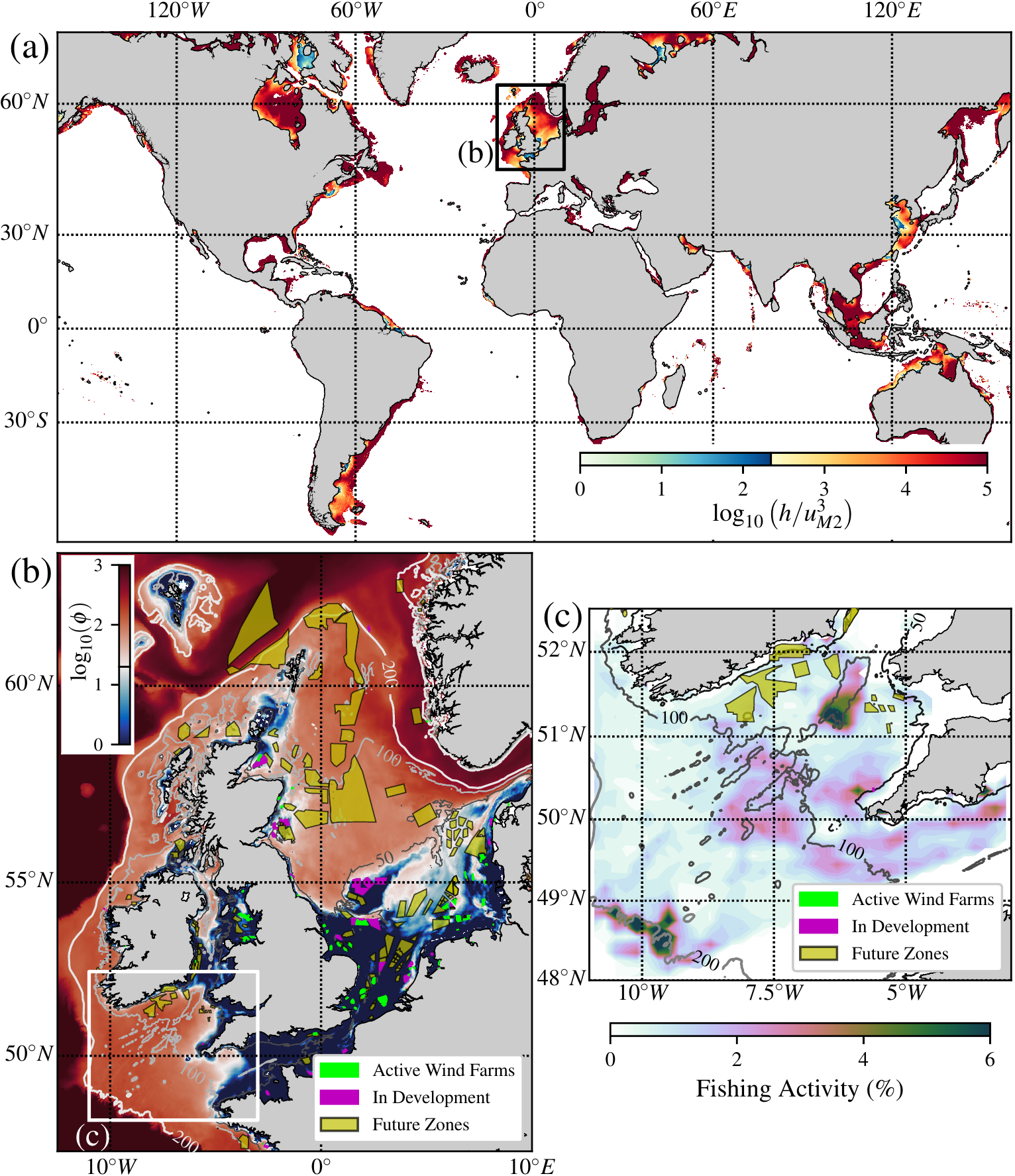}};
\draw[draw=black,fill=white] (-7.6,0.9) rectangle ++(3.9,4.9);
\node at (-5.8,3.2) {%
  \begin{tabular}{lcc}
  \multicolumn{1}{c}{Region}  & \begin{tabular}[c]{@{}l@{}}2020\\ (GW)\end{tabular} & \begin{tabular}[c]{@{}l@{}}2030\\ (GW)\end{tabular} \\
  \hline  \\[0.1cm]
  Europe & 25 & 114 \\[0.3cm]
  \begin{tabular}[c]{@{}l@{}}North\\ America\end{tabular} & \multicolumn{1}{c}{0} & \multicolumn{1}{c}{24} \\[0.3cm]
  China & 10 & 64 \\[0.3cm]
  \begin{tabular}[c]{@{}l@{}}Asia (Ex.\\ China)\end{tabular}  & \multicolumn{1}{c}{0} & \multicolumn{1}{c}{36} \\[0.3cm]
  Other & 0 & 5 \\
  \end{tabular}
};
\end{tikzpicture}
\caption{
Offshore wind and seasonally stratified shelf seas. 
(a) Shelf sea development; the limited extent of well-mixed waters, defined using the ``$h/u_{M2}^3$'' criterion, is highlighted against projected global offshore wind energy growth \citep{gwec2020global}.
(b) NEMO model of the Northwest Europe summer potential energy anomaly, $\phi$, a measure of the amount of stratification \citep{guihou2018kilometric}.
(c) Fishing hotspots in the Celtic sea caused by topographically-enhanced mixing of stratified waters \citep{sharples2013fishing}.
In (a) and (b) seas are partitioned into regions prone to seasonal stratification and those remaining well mixed, based on: (a) $h/u_{M2}^3= \SI{220}{\second \cubed \per \meter \squared}$  \citep{simpson2012introduction}; and (b) $\phi = \SI{20}{\joule \per \meter \cubed}$ \citep{gowen1995regional}.
In (b) and (c) offshore wind farms are separated into active sites, sites still in construction or development and identified zones for future development \citep{4coffshore}.
}
\label{fig:globalwind}
\end{figure}
Renewable energy solutions, including offshore wind, are prerequisite for clean growth and thus the reduction of greenhouse gas emissions needed to mitigate against climate change. 
Offshore wind energy in shelf seas has seen a rapid increase over the past decade \citep{xu2020proliferation,diaz2020review}, motivated by: high-quality and reliable energy (wind) resources \citep{esteban2011offshore}; space availability and site accessibility for installation of large, efficient, turbine systems \citep{sun2012current}; rapidly maturing, reliable and energy-efficient technologies \citep{jansen2020offshore}; and reduced visual impact on populated areas \citep{wen2018valuing}. 
Government programmes have helped drive development of renewable offshore wind energy from offshore wind farm arrays, of tens increasing to hundreds, of offshore wind turbines (OWT) supported by various fixed foundation designs with new floating foundations being designed to access deeper water sites.
Northwest Europe has led sector development, with the UK leading in GW operational capacity to date \citep{gwec2020global}.
The sector has grown rapidly, with technological advances reducing the Levelised Cost of Electricity to a point where  price is competitive with alternative energy solutions \citep{shen2020comprehensive}. 
Thus, to meet demand, global development of offshore wind energy in shelf seas is predicted to grow from 35 GW operational in 2020 to 243 GW operational by 2030 (Figure \ref{fig:globalwind}a).
\par
With over 80\% of the global population living within 100 km of the ocean, shelf seas have significant economic and social value, including fishing, shipping, carbon storage (``blue'' carbon) and recreation.  
Despite comprising 8\% of the total area of the global ocean (Figure \ref{fig:globalwind}a), shelf seas support 15 – 30\% of global ocean biological production \citep{wollast1998evaluation}.  
This biological production ultimately supports $>90\%$ of the world’s fish landings \citep{pauly2002towards} and plays a disproportionately important role in the absorption of CO$_2$ from the atmosphere \citep{roobaert2019spatiotemporal}. 
Thus, high biological productivity means shelf seas are key components of global biogeochemical cycles, supporting societally important bioresources and also the biological uptake and storage of carbon in the marine environment. 
However, interplay of social and economic drivers already places significant stress on shelf seas \citep{kroger2018SSBP}. 
Further industrialization of shelf seas will enhance these stresses, with the potential for significant long term environmental impact.
Shelf sea dynamics directly control primary production: the growth of microscopic marine plankton. However, from OWT scale to coastal scale, the impact of offshore wind development on shelf seas has yet to be fully considered.
Therefore, future offshore wind energy development must be grounded in advanced understanding of impact on shelf sea dynamics.
This is critical to enable balance of key global Societal Goals, i.e. to “ensure access to affordable, reliable, sustainable and modern energy” and to “conserve and sustainably use the oceans, seas and marine resources” \citep{UNSDG}.
\par
\begin{figure}
\centering
\includegraphics[width=\textwidth]{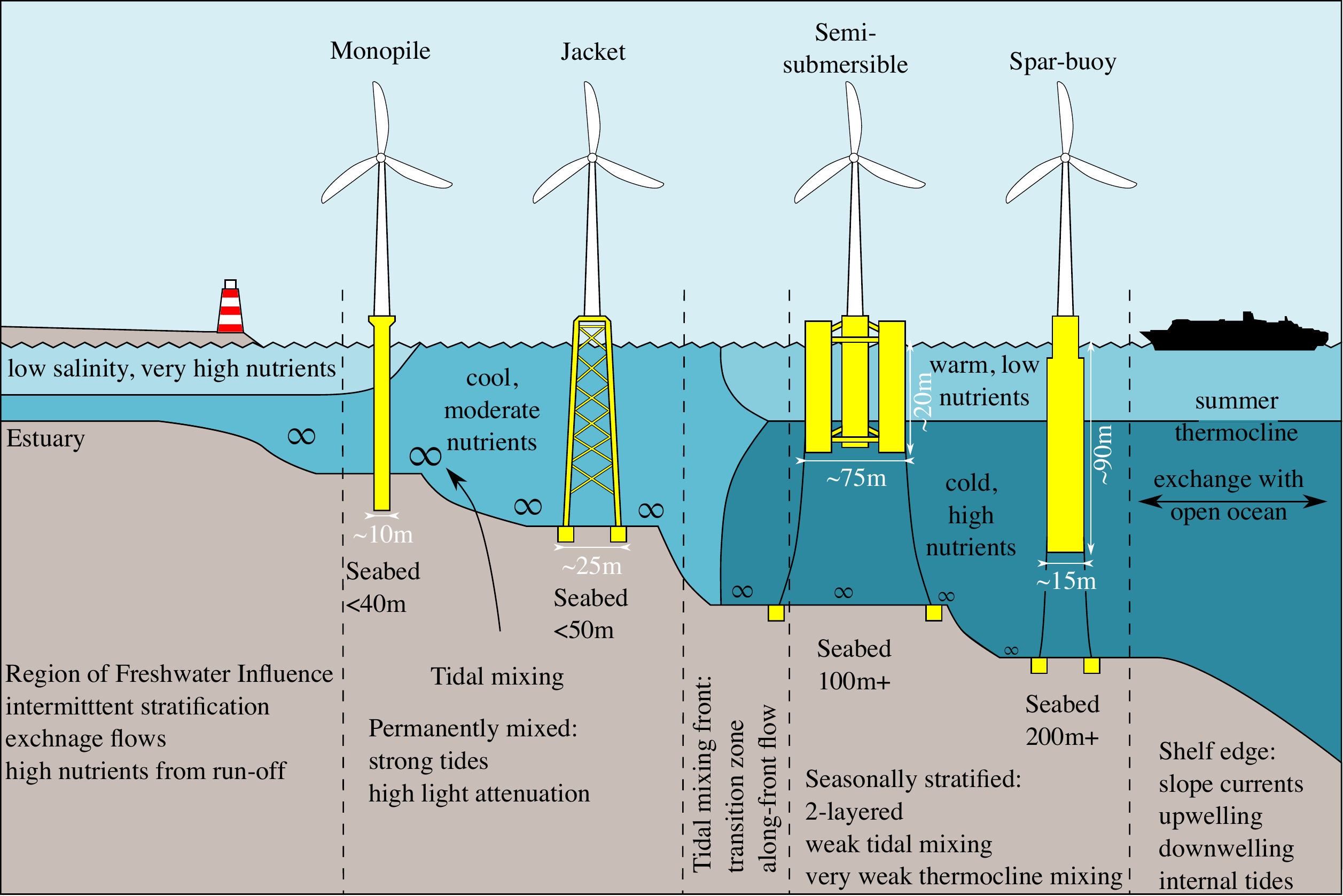}
\caption{Existing and emerging offshore wind engineering solutions \citep{diaz2020review,jiang2021installation}, including fixed monopile and jacket foundations and floating semi-submersible and spar-buoy foundations, in comparison to shelf sea regimes from coastline to open ocean \citep{simpson2012introduction}.
}
\label{fig:structures}
\end{figure}
To date, most offshore wind farms have been installed in the near-shore shallow water regions, up to 50 m depth, of shelf-seas (Figure \ref{fig:globalwind}b).
Near-shore shallow-water installations have been preferred due to the cost reduction from ease of access for installation, grid connection and operation and maintenance \citep{jacobsen2019nearshore}.
With sector plans for an additional 208 GW of operational capacity in the next decade, and targets of 1.4 TW total by 2050 \citep{OREAC20}, near-shore and shallow-water sites are rapidly becoming limited.
The scale of expansion of offshore wind energy means the sector is now expanding into deeper water sites further from shore \citep{soares2020current}.
The transition from near-shore and shallow-water environments to deeper water further from shore marks a fundamental change in the marine environment. 
Shallow waters are typically well-mixed; however deeper waters may be subject to seasonal stratification, where density varies vertically with depth (Figure \ref{fig:globalwind}a and b). 
Stratified waters are a vital part of shelf seas, controlling primary production and biogeochemical cycling \citep{simpson2012introduction}.
Expansion into this new environment means that offshore wind farms will increasingly come into conflict with its environmental functioning, controlled by natural mixing of water column stratification (Figure \ref{fig:globalwind}c).
\par
Addressing engineering challenges, both fixed and floating foundations are being developed to enable expansion into deeper waters.
Fixed foundations, which span the entire water depth, include monopiles, gravity bases and jacket constructions \citep[see, e.g, Figure \ref{fig:structures} and ][]{esteban2019gravity,diaz2020review,jiang2021installation}.
However, floating foundations are crucial to deep water, $>50$ metres, deployment.
Learning from the petroleum industry \citep{schneider2010foundation}, designs include tension-leg platforms \citep{uzunoglu2020hydrodynamic}, spar-pendulum \citep{cottura2021dynamic} and spar-buoy platforms \citep{jacobsen2021influence}, and semi-submersible platforms \citep{castro2020economic}.
Using the submerged structural buoyancy and mooring forces to balance atmospheric thrust and wave loads, floating foundations typically have large draft, e.g. spar platforms, or large cross sectional area, e.g. semi-submersible platforms \citep{butterfield2007engineering}.
Thus, with sector development requiring larger of turbines that need bigger rotors, which are subject to greater atmospheric loads, the draft and diameter of fixed and floating foundations will need to increase.
However, the $>20$ m draft of current floating foundations is already large enough to penetrate the thermocline and directly mix seasonally stratified shelf seas (Figure \ref{fig:structures}).
\par
For the first time, large scale industrialisation of seasonally stratified marine environments is planned.
Over two decades of research has already focused on the direct impacts of offshore wind farm development on well-mixed shallow water marine ecosystems, from: benthic habitats \citep{dannheim2020benthic}, fisheries \citep{gray2005offshore} to seabirds \citep{exo2003birds}.
Whilst this research is translatable with sector growth, the seasonally stratified regime offers a fundamentally new challenge: the introduction of infrastructure will lead to enhanced `\textit{anthropogenic}' mixing of stratified waters. 
Enhanced mixing may lead to profound impacts on shelf sea dynamics and thus marine ecosystem functioning.
The aim of this paper is to investigate the scope of these potential impacts. 
Section 2 reviews the ecosystem and physical functioning of stratified shelf seas. 
Section 3 then describes our current understanding of the impact of offshore infrastructure on unstratified and stratified waters.
Section 4 discusses current research challenges, the potential impact of offshore wind on stratified shelf seas and the sector requirements needed to ensure acceleration of renewable energy and its sustainable development.
It is concluded that offshore wind farm infrastructure may have significant, and long lasting, effects on fragile shelf sea ecosystems.
Criteria for Environmental Impact Assessments, must therefore be revised and updated to enable the sustainable growth, and acceleration, of renewable offshore wind energy development.

\section{Oceanography of Stratification in Shelf Seas}
\label{section2}
Shelf seas lie on the continental shelf between the coast and the continental slope, where at 200 m water depth the sea floor slopes down to the deep ocean. 
Despite only accounting for 0.5$\%$ of ocean volume shelf seas play a key role in the Earth system, dissipating $>$ 70$\%$ of the tidal energy \citep{egbert2000significant} and are disproportionately important in supporting ocean biological production\citep{wollast1998evaluation}, fish landings \citep{pauly2002towards} and the absorption of CO$_2$ from the atmosphere \citep{roobaert2019spatiotemporal}. 
Biological production is underpinned by the growth of microscopic marine phytoplankton, which is tightly controlled by the timing and strength of seasonal stratification.    
The seasonally stratified zones in shelf seas act as an important net sink of carbon \citep{thomas2004enhanced}. 
This makes the physical and biogeochemical processes described here a key dynamic component of the global carbon cycle \citep{bauer2013changing} linking the atmospheric, terrestrial, and oceanic carbon pools.
\begin{figure}
\centering
\includegraphics[width=\textwidth,trim={0 8cm 0 0},clip]{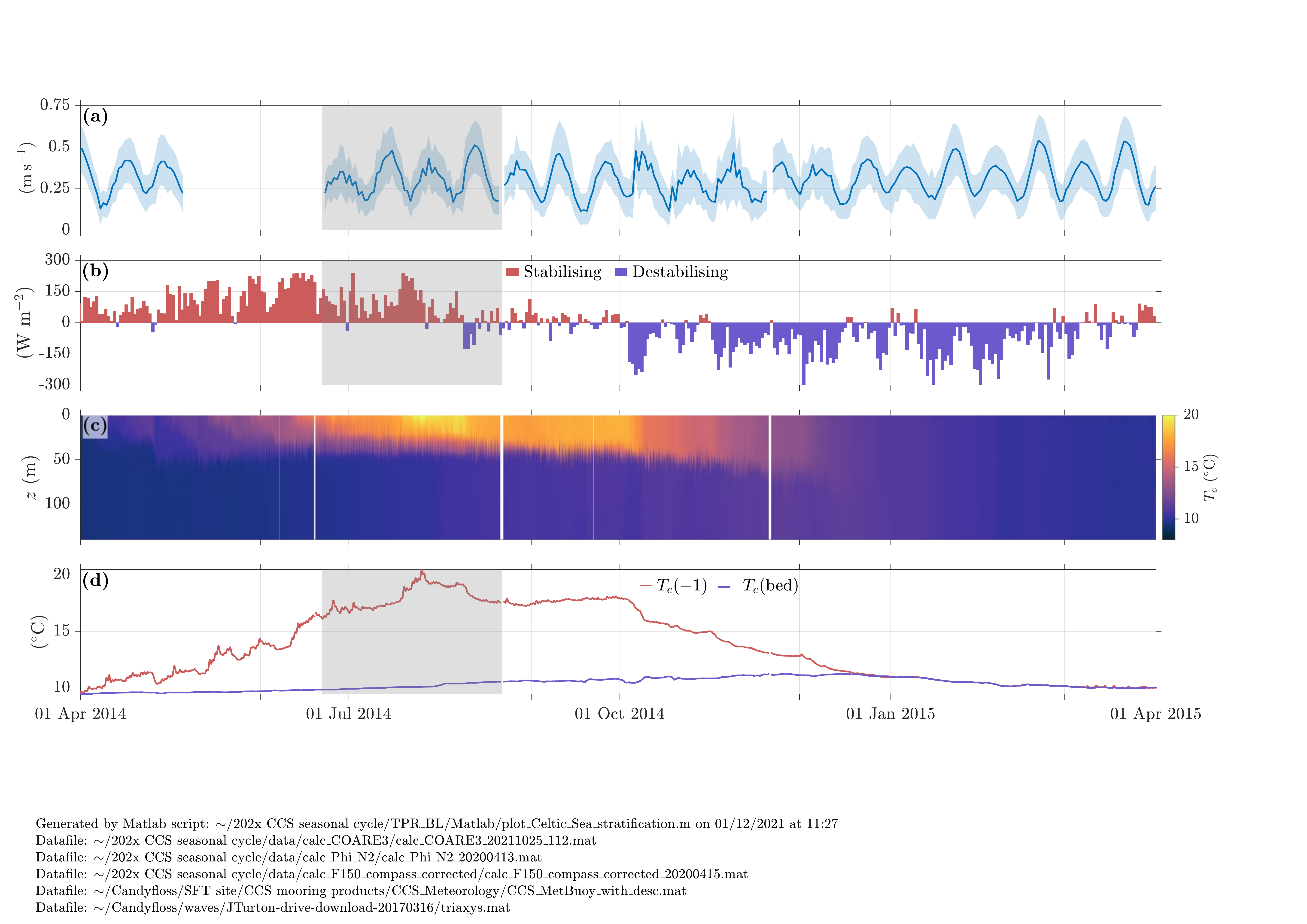}
\caption{
Seasonal time-series of mooring observations from 2014 in the Celtic Sea ($49^{\circ} 24$ N $8^{\circ}  36$ W), collected as part of the UK Natural Environment Research Council (NERC) funded CaNDyFloSS project (NE/K00168X/1) \citep{scannellthesis}. 
a) Measured thermocline current speed over 20-50 m mean depth; average speed  (blue line), and range (shaded region) over 2 tidal periods. 
b) Surface heat exchange. c) Vertical temperature structure. d) Surface and bottom layer temperatures.}
\label{fig:seasonal timeseries}
\end{figure}
\newline
\subsection{Distribution and Seasonal Cycle of Stratification}
Away from coastal regions of freshwater influence, the water column structure in temperate shelf seas undergoes a seasonal cycle in response to changes in heat exchange at the surface.
In spring and summer, some areas of the temperate shelf seas become thermally stratified whilst neighbouring areas remain well mixed.  
The first order control on the water column structure has been established as a balance between the stratifying influence of surface heating and turbulent mixing due to the tides \citep{simpson1974fronts}.  
In regions of shallow water and strong tidal currents (of order metres per second), the rate of buoyancy input due to surface heating is insufficient for the establishment of persistent stratification, and in consequence the water column remains homogeneous. 
However, in regions of deeper water and/or weaker tidal currents (and associated lower levels of turbulence), surface heating dominates and seasonal stratification develops.
Away from coastal regions, shelf seas generally exhibit tidal currents of order of tens of cm per second, meaning that stratification are typically found when water depths are greater than 80 m.  
In stratified regions, a warmer surface water layer 5-40 m thick overlies a deeper cooler water layer. 
The two layers are separated by a region of strong vertical temperature gradient, the thermocline, which forms a barrier to vertical exchange of heat, salt, nutrients and momentum. 
\par 
Tidal mixing fronts separate regions of seasonal stratification from well mixed regions. \citet{simpson1974fronts} use an energetics argument to derive a single parameter to predict the positions of these fronts. 
By considering only vertical exchange processes and assuming the surface input of heat was the only stratifying influence, and that tidal currents are the only source of energy driving mixing, they showed that the first order determinate for the position of shelf sea fronts is given by the ratio $h/u_{M2}^3$, where $u_{M2}^3$ is the principle lunar M2 tidal current amplitude, and $h$ is water depth  (Figure \ref{fig:globalwind}a). In Figure (\ref{fig:globalwind}a) $h/u_{M2}^3$ is calculated from bathymetry and M$_2$ tide data taken from TPX09 global tidal atlas \citep{egbert2002efficient}, which applies a generalised inverse method, assimilating satellite altimeter data, into a global barotropic tidal model; here model resolution limits precise location of stratified fronts from global data.
In terms of area, regions of seasonal stratification dominate the continental shelf seas (Figure \ref{fig:globalwind}a).
Whilst the critical value for the ratio characterising the position of tidal mixing fronts was initially estimated for the Irish Sea \citep{simpson1974fronts,simpson1981models}, consistent values have subsequently been estimated for a range of shelf seas globally [e.g. the Gulf of Maine and Bay of Fundy, \citet{garrett1978tidal,loder1986predicted}; the Yellow Sea, \citet{lie1989tidal}; the Patagonian Shelf, \citet{glorioso1995barotropic}; the northwest European Shelf Seas, \citet{pingree1978tidal} and the Bering Sea, \citet{schumacher1979structural}]. 
The robustness of the critical value highlights the key role of the tides in determining the position of shelf sea fronts and provided the first quantitative link between the dissipation of tidal energy and ocean mixing.
\par
The strength of stratification may be quantified in terms of the potential energy anomaly, $\phi$ (\SI{}{\joule \per \m \cubed}), which describes the energy required to fully mix a stratified water column (Simpson and Bowers, 1981) 
\begin{equation}
 \phi =\frac{g}{h}\int_{-h}^{0}(\overline{\rho} -\rho  )z{\rm d}z,
\end{equation}
where $h$ is the water depth and $\rho$ the water density, $\overline{\rho}$ denotes the density calculated using the mean water temperature and salinity \citep{holt2008seasonal}. Geographical variation of $\phi$ across the NW European Shelf Seas is plotted on the map in Figure \ref{fig:globalwind}b.
\par
A typical seasonal stratification cycle shows a time-series of warming and cooling, varying with water depth (Figure \ref{fig:seasonal timeseries}). 
Surface mixed layer temperatures warm from April into the summer in response to a net positive buoyancy input due to surface heating.  
The observed surface temperatures of $15-20$ \SI{}{\celsius} are typical for temperate shelf seas in midsummer, while the deep water remains close to its winter temperature of \SI{10}{\celsius}.
Over this period the strength of stratification grows with a surface to bed temperature difference exceeding \SI{10}{\celsius} by August, which gives rise to a density difference of \SI{0.3}{\kg \per \m \cubed}. 
The stratification weakens into autumn as surface cooling leads to convection and storms drive turbulent mixing, such that the water column becomes well mixed during the winter.
\par
During the stratified period the deeper water is isolated from the surface layer by a thermocline.
A slow warming of the deep-water results from mixing down of heat from the sea surface.
The rate of warming, set by thermocline mixing,  varies geographically and is important as it determines the timing of the autumnal breakdown of stratification \cite{rippeth2005mixing}, and transport of nutrient rich deep water up to the surface layer.
\newline
\subsection{Ecosystem Response to Stratification}
Primary production of organic matter by phytoplankton forms, directly or indirectly, the primary food source for almost all marine organisms.
Phytoplankton growth requires CO$_2$, sunlight and nutrients, the availability of which are determined by water column structure, with profound implications for the biological functioning of the shelf seas.  
\begin{figure}
\centering
\includegraphics[width=\textwidth]{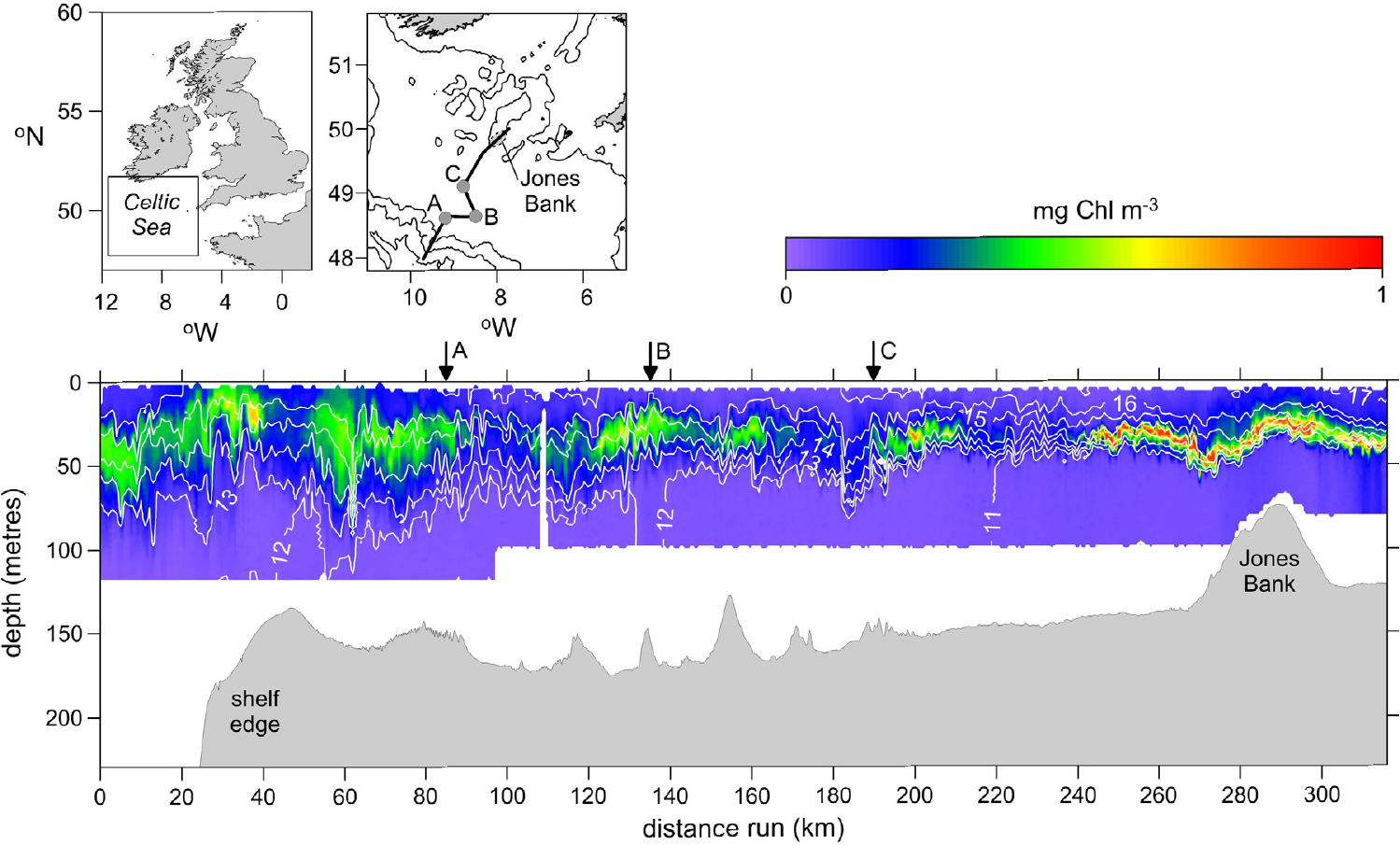}
\caption{Observational data from the Celtic Sea collected in summer 2008, supported by NERC's Oceans2025 Programme \citep{sharples2013fishing}.  Section of temperature (line contours) and chlorophyll concentration (colours) measured using a Scanfish CTD towed along the path show on the map.  High chlorophyll concentration in the SCM (subsurface chlorophyll maximum) indicate phytoplankton production extending hundreds of kms across the shelf.
Enhancements in concentration over rough topography, such as Jones Bank, are a result of elevated turbulent mixing, driving nutrient fluxes which correspond directly to hotspots of marine biodiversity and thus fisheries (Figure \ref{fig:globalwind}c).}
 \label{fig:SCMprofile}
\end{figure}
\par
In turbulent unstratified regions, primary production occurs mainly during summer months when sunlight is strong.  
However, plankton are continuously mixed from sea surface to bed by turbulence, and spend much of their time below a depth where light intensity is sufficient for growth. 
In contrast, stratified waters provide ideal conditions for phytoplankton growth in spring.
As stratification forms phytoplankton become trapped in the well lit surface layer, with the thermocline acting as a barrier to mixing.  
Phytoplankton retained in the surface layer enjoy the abundance of light, and exhibit rapid growth forming the annual ``spring bloom'', a biological abundance visible from space that forms the year’s first supply of significant new organic fuel.  
As the phytoplankton grow, they fix inorganic carbon in the surface water into organic carbon, which causes the sea surface to replenish its dissolved carbon concentration by absorbing CO$_2$ from the atmosphere. 
The timing of the spring bloom is so significant that zooplankton and fish larvae have evolved to use it as a food source \citep{platt2003spring}, with further implications higher up the food-web, e.g. for shrimp survival \citep{ouellet2011ocean} and seabird breeding success \citep{frederiksen2006plankton}. 
\par
During the spring bloom, the availability of nutrients in the surface layer becomes exhausted, and further production is limited by nutrient supply.  
Despite this limitation on plankton growth, a persistent and significant level of primary production is sustained at depth, throughout the period of seasonal stratification.  
This sub-surface phytoplankton layer located in the stratified thermocline water is a ubiquitous feature and is known as the ``subsurface chlorophyll maximum'' (SCM)  \citep{pingree1982celtic}.  
In shelf seas, the SCM occupies a 10-30 m thick layer, the depth of which varies from 10-40 m across the hundreds of kms that it extends over the shelf (Figure \ref{fig:SCMprofile}).
The SCM plays a vital role in supporting the pelagic food web during summer. 
Estimates based on observations of primary production rates within the SCM suggest that subsurface carbon fixation accounts for up to $50\%$ of annual primary production in the seasonally stratified North Sea \citep{richardson2000subsurface,weston2005primary}. 
An extrapolation using microstructure-based nitrate flux estimates also gives the same approximate figure \citep{rippeth2009diapcynal}.
The persistence of production in the SCM is dependent on a vertical flux of nutrients from the deep nutrient-rich water below \citep{sharples1994modelling}.  
In consequence, the processes responsible for mixing across the thermocline, discussed in section \ref{sec:small}, are key to delivering the limiting nutrients to the euphotic zone and sustaining the SCM \citep{sharples2001phytoplankton,sharples2001internal,sharples2007spring,williams2013maintenance}.  
Episodic mixing associated with storm initiated inertial oscillations \citep{burchard2009generation,lincoln2016surface} have been shown to drive significantly enhanced nutrient fluxes \citep{williams2013wind}.
Local enhancement of primary productivity is evident in regions of steep topography, where tidally induced internal waves elevate mixing and nutrient fluxes.  
Chlorophyll concentration in the SCM is greatly elevated at the shelf break \citep{sharples2007spring}, and mid-shelf sand banks \citep{sharples2013fishing}.  
The dependence of the marine food web and fisheries on upward flux of nutrients is evident in the distribution of seasonal fishing hot-spots (Figure \ref{fig:globalwind}c).
\par
The organic products of the spring and summer primary production sink into the deeper waters, where bacteria remineralise the organic material (nutrients and carbon) back to the inorganic components. 
Remineralisation removes oxygen from the deep water.
In addition, the barrier role of the thermocline limits the replenishment of that oxygen from the atmosphere \citep{mahaffey2020impacts}.
Together both processes determine the dissolved oxygen concentrations available to benthic and pelagic organisms.
Thus, high shelf sea biological productivity means shelf seas are key components of global biogeochemical cycles, supporting societally important bioresources, and also the biological uptake and storage of carbon in the marine environment.
\newline
\newline
\subsection{Shelf Sea Mixing Processes}
\label{sec:seamix}
Currents in shelf seas provide energy for stirring the water column and are generally dominated by tidal motions, with episodic contributions by the wind in the upper water column.
An example of the current variability from the Celtic Sea, a typical shelf sea location, is presented in Figure \ref{fig:seasonal timeseries}a, where velocities vary from 0.1-0.7 \SI{}{\m \per \second}.
For this location, the semi-diurnal tide lunar M$_2$ produces two high and low tides a day with 4 peaks in current speed.
The interaction with the principle solar tidal component, S$_2$, produces the 14 day spring-neap cycle.
Wind driven currents are also observed in the top 50 m, and take the form of inertial oscillations, which have a latitude dependent period, which is 14.9 hours at the mooring location.
\begin{figure}
\centering
\includegraphics[scale=1]{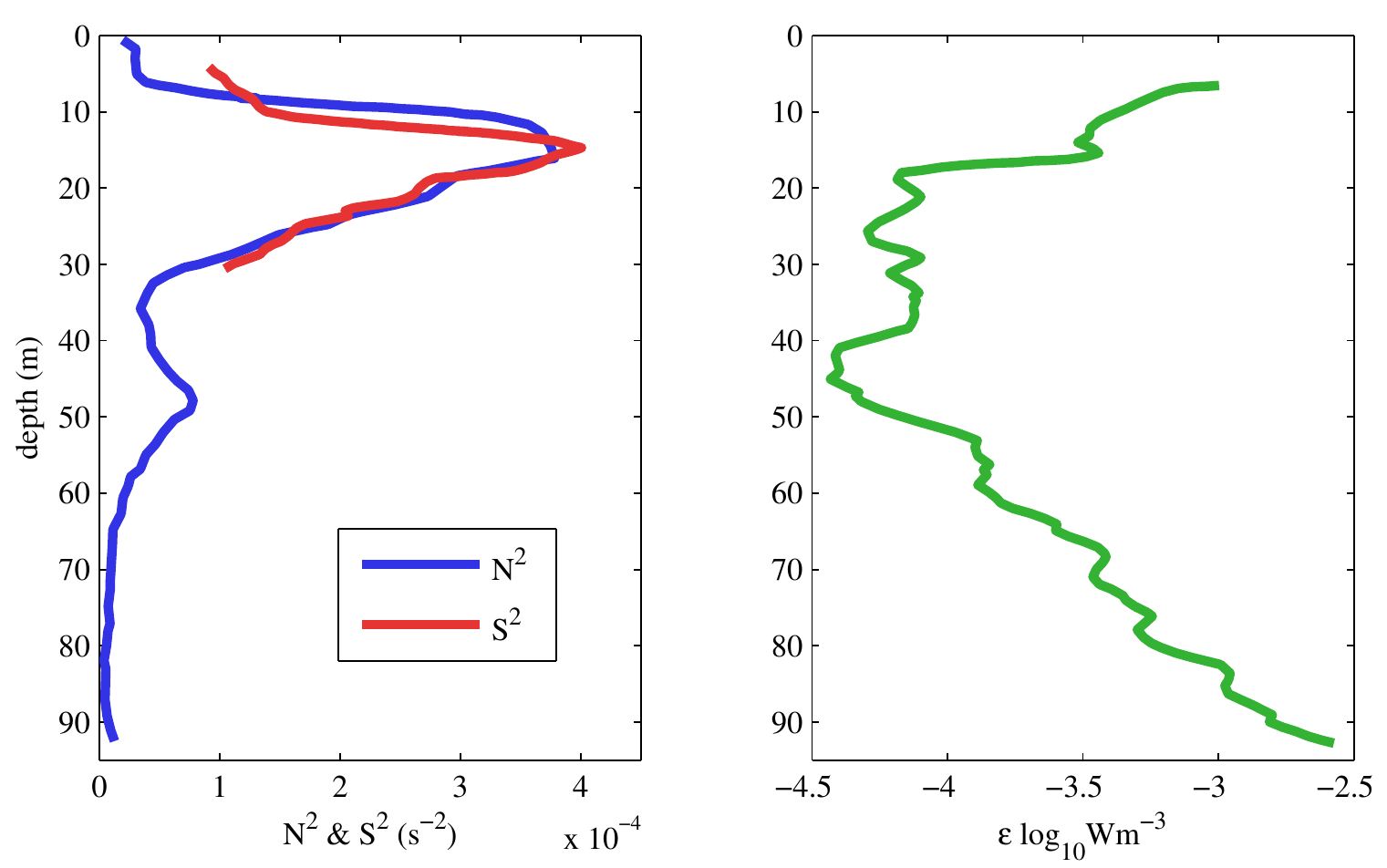}
\caption{Marginal stability and energy dissipation in seasonally stratified shelf seas. 
a) The equivalence between buoyancy frequency $N^2$ (blue) and vertical current shear $S^2$ (red), temporally averaged (over two tidal cycles). 
b) The coincident temporally averaged profile of TKE dissipation rate $\varepsilon$.
Measurements from direct observations in the Western Irish Sea in June 2002 \citep{rippeth2005mixing}.}
\label{fig:marginal}
\end{figure}
\par
Friction at the seabed and the sea surface generate vertical current shear in the flow, and turbulent eddies which “cascade” to ever smaller scales until their energy is dissipated either to heat, or to potential energy via mixing.
\par
In the absence of convection, a three-way local balance is assumed (assuming both at least quasi-stationarity and that transport terms can be ignored):
\begin{equation}
    \label{eq:prod}
    \mathcal{P}  = \mathcal{B} + \varepsilon,
\end{equation}
where $\mathcal{P}$ is the (total) production of turbulent kinetic energy (TKE), $\mathcal{B}$ is the buoyancy production (mixing), and $\varepsilon$ is the TKE dissipation rate (heat). 
\par
The efficiency of mixing by turbulence can be quantified by the flux Richardson number $Rf = \mathcal{B}/\mathcal{P}$ and is widely assumed to have a value $Rf \ll 1$ in a stratified fluid. Since $\varepsilon$ is a commonly measured turbulence metric it is often used to infer the rate of mixing using the closely related flux dissipation coefficient, defined in terms of the buoyancy production as $\Gamma = \mathcal{B}/\varepsilon$.   
A value of $\Gamma \approx 0.2$ (i.e. $Rf \approx 1/6$) is routinely applied, and has been verified for the shelf sea thermocline by a number of different observational approaches \citep{oakey2004mixing,inall2000impact,palmer2008investigation,bluteau2013turbulent}, though it has been found to vary in other regimes \citep{monismith2018mixing}.
\par
A consequence of shear production at the seabed by barotropic tidal currents is that measured rates of turbulence are extremely low in the seasonal thermocline, orders of magnitude lower than at the boundaries (Figure \ref{fig:marginal}b). Mean dissipation rates range from $\varepsilon=10^{-7}  - 10^{-5} \ \SI{}{\watt \per \m \cubed}$ \citep{rippeth2005mixing}, 2-3 orders of magnitude smaller than rates commonly found in shallow well mixed waters \citep{simpson1996vertical}. Empirical estimates of the bulk mixing efficiency of the barotropic tide in stratified waters, are very low, $Rf_{BT}\approx0.0037$ \citep{simpson1978fronts}, as most turbulence is produced in the well mixed bottom layer, so that no mixing is possible.   
In addition, strong density gradients in the thermocline inhibit vertical mixing, and as such rates of vertical mixing observed in shelf seas are comparable to using a hand mixer in a swimming pool.
\par
The production of turbulence in stratified waters is inhibited by buoyancy forces arising from vertical density gradients, which are quantified by the Brunt-V\"ais\"al\"a, or buoyancy, frequency, $N$, where
\begin{equation}\label{eq:N}
    N = \sqrt{-\frac{g}{\rho}\frac{\partial \rho}{\partial z}},
\end{equation}
describes the frequency at which a displaced parcel of fluid will oscillate in a stratified system and is thus a measure of the stability of stratified waters.
Conversely, the vertical current shear,  $S = \partial u / \partial z$, is a measure of the extraction of energy from the mean flow, and therefore power available to overcome buoyancy forces and generate turbulence.
The generation of instabilities in stratified water is quantified using 
measurements of the buoyancy frequency, $N$, and vertical current shear, $S$, to calculate the gradient Richardson Number
\begin{equation}
    Ri_g = -\frac{g}{\rho}\frac{\frac{\partial \rho}{\partial z}}{\left(\frac{\partial u}{\partial z}\right)^2} \equiv \frac{N^2}{S^2}.
\end{equation}
Both theory \citep{miles1961stability,howard1961note} and observations \citep{silvester2014observations,lincoln2016surface} show that shear instability and internal wave breaking occurs when $Ri_g \lesssim 0.25$.   
Average of measurements, over two tidal cycles, show $N^2$ and $S^2$ are approximately equal implying a (close to) marginally stable water column where $Ri_g \approx 1$ (Figure \ref{fig:marginal}a).
Therefore, any additional sources of shear can generate instabilities and mixing.  
This marginal stability has been widely observed across shelf seas \citep{van1999strong,rippeth2005mixing,palmer2008investigation} but it does not universally explain the observations \citep{Palmer2013} suggesting additional sources of shear may, at times, control system dynamics.
\par
One dimensional vertical exchange models using turbulence closure schemes (also commonly referred to as RANS models) fail to represent the measured rates of mixing through tidal and wind driven boundary processes \citep{simpson1996vertical}.
The deficit in the predicted mid-water $\varepsilon$ points to either an incorrect parameterisation of the small scale physics away from the boundaries or to the absence of key physical processes in the model. 
A recent study \citep{Luneva19} evaluated a range of alternative one dimensional turbulence closure schemes for the Northwest European shelf seas, packaged within the Generic Length Scale two-equation formulation \citep{UmlaufBurchard05}. 
Evaluating the  schemes against profile data (28000 profile in total) confirmed that there was no outright winner, with all schemes under representing the thermocline properties and suggested that physical processes are still missing. 
Candidate mechanisms to account for the deficit in mid water mixing include internal waves generated by stratified flow over topography and wind generated inertial currents.
\par
Internal tides propagate at the thermocline in response to tidal currents flowing over steep topography \citep{rippeth2005mixing,inall2021shelf}.
These waves generate strongly sheared currents about the thermocline which can lead to the development of shear instability, draining energy to turbulence which then supports mixing in the quiescent mid-water region.  
Although the energy in the internal tide is much less than the barotropic tide, an empirical estimate of bulk mixing $Rf_{IT}\approx0.056$ \citep{stigebrandt1989vertical} shows it to be highly efficient compared with the barotropic tide $Rf_{BT}\approx0.0037$ \citep{simpson1978fronts}, on account of the turbulence coinciding with the stratification. A consequence of the differing bulk mixing efficiencies is that whilst the barotropic tide dominates the budget for the dissipation of barotropic tidal energy in the seasonally stratified shelf seas to the west of the UK, the shelf break generated internal tide is estimated to dominate diapcynal mixing \citep{rippeth2005thermocline}. 
With advances in computational performance and associated resolution increase, simulation of internal tides are now routine for European Shelf simulations \citep{guihou2018kilometric} implemented operationally \citep{Graham2018, Tonani2019}. 
Though, as hydrostatic simulations, the non-hydrostatic mixing processes must still be parameterised.
\newline
\subsection{Submesoscale Mixing}
\begin{figure}
\centering
\includegraphics[width=\textwidth]{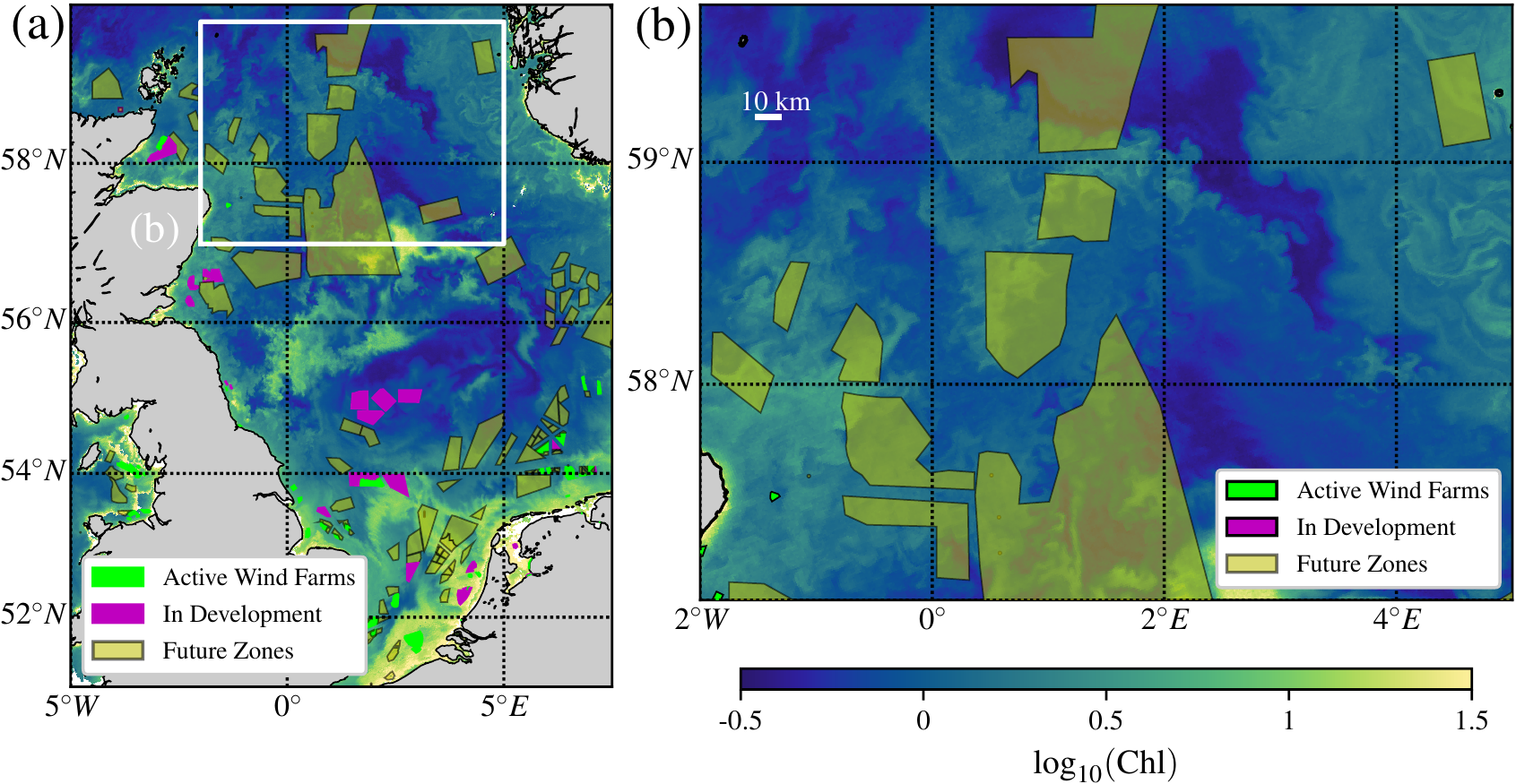}
\caption{Near-surface chlorophyll concentration (\SI{}{\mg \per \m \cubed}) from 20
April, 2020, inferred from the NASA MODIS Aqua satellite. 
a) Chlorophyll concentration for Northwestern European waters. 
b) Detail of the north North sea highlighting submesoscale instabilities at, and below, the scale of offshore wind farm development. 
The image was made from data downloaded from \url{http://oceancolor.gfsc.nasa.gov} and \citet{4coffshore}.}
\label{fig:chl_submesoscale}
\end{figure}

The size of many existing and proposed offshore wind farms coincides with the ocean submesoscale ($\sim 1-20$ km). The imprint of submesoscale eddies can be seen in satellite images of chlorophyll concentration in the North Sea as seen in Figure \ref{fig:chl_submesoscale} (note that the future development zones shown for reference will likely consist of multiple smaller windfarms). 
The submesoscale range is characterised by Rossby numbers, $Ro\equiv u/(fL)\sim 1$, where $u$ and $L$ are characteristic horizontal velocity and length scales associated with submesoscales and $f$ is the Coriolis parameter. 
This makes submesocales dynamically distinct from mesoscales ($\sim 100$ km) where $Ro\ll 1$. While the Earth's rotation is important on submesoscales, it does not constrain the flow as strongly as it does for larger scales, leading to a unique set of physical processes. 
Submesoscale currents are generally more energetic in regions with large horizontal density contrasts which includes coastal waters. 
\par
It is not clear if, or how, offshore wind farms might interact with submesoscale currents. 
However, since the spatial extent of wind farms typically lie within the submesoscale range, array scale flow patterns generated by enhanced drag or mixing within offshore wind farms could be subject to the physical processes that are active on the submesoscale. 
Below we very briefly consider some of these processes. 
See \citet{thomas2008submesoscale,mcwilliams2016submesoscale,meredith_book_submesoscale} for in-depth reviews of submesoscale dynamics.
\par
Submesoscales are typically generated by dynamical instabilities or flow/topography interactions. 
The instability mechanisms include ageostrophic versions of instabilities that also exist at larger scales (e.g.\ baroclinic instability) and instabilities that are unique to the submesoscale (e.g.\ inertial/centrifugal and symmetric instability). 
Although the details and energy pathways differ, the net effect of submesoscale instabilities is to convert potential energy associated with horizontal density gradients into kinetic energy. 
In the process, submesoscales increase the water column stratification and tend to reduce the depth of the surface mixed layer. Submesoscales can also be generated as currents move around topographic features. For example, submesoscale eddies can be generated by flow past islands \citep{marmorino2018}, along continental slopes \citep{molemaker2015,gula2015}, and over seamounts \citep{srinivasan2019}. It is possible that the enhanced drag experienced by the flow through a submesoscale offshore wind farm could similarly generate submesoscale eddies. The enhanced mixing and drag within wind farms could also influence submesoscale instabilities and eddies, although these hypotheses remain untested.
\par
Submesoscales are important partly because they interact with vertical mixing processes. Submesoscales are characterised by timescales that range from hours to $\sim 1$ day. 
As a result, the tendency for submesoscales to increase the vertical stratification of the water column is fast enough to compete with vertical mixing driven by winds and tides. 
At the same time, submesoscales can induce strong vertical circulations and locally enhance the exchange between the surface mixed layer and thermocline \citep{mahadevantandon}. \citet{gula2015} provide an in-depth review of the connections between submesoscales and ocean mixing.
\par
Submesoscales can have a strong impact on biogeochemistry \citep{levy2012,mahadevan2016}. 
For example, shoaling of the surface mixed layer depth and suppression of turbulent mixing induced by submesoscales can trigger phytoplankton blooms in otherwise light-limited conditions \citep{taylorferrari_GRL,mahadevan2012,taylor2016}. 
In nutrient-limited conditions, submesoscales can upwell nutrient-rich waters to the euphotic zone, enhancing primary production \citep{levy2001,mahadevanarcher}. 
Finally, downwelling circulation and the suppression of turbulent mixing can enhance the export of particulate organic matter from the surface mixed layer \citep{omand2015,taylor2020}. 
\newline
\subsection{Mixing Across Density Interfaces}
\label{sec:small}
\begin{figure}
\centering
\includegraphics[width=\textwidth]{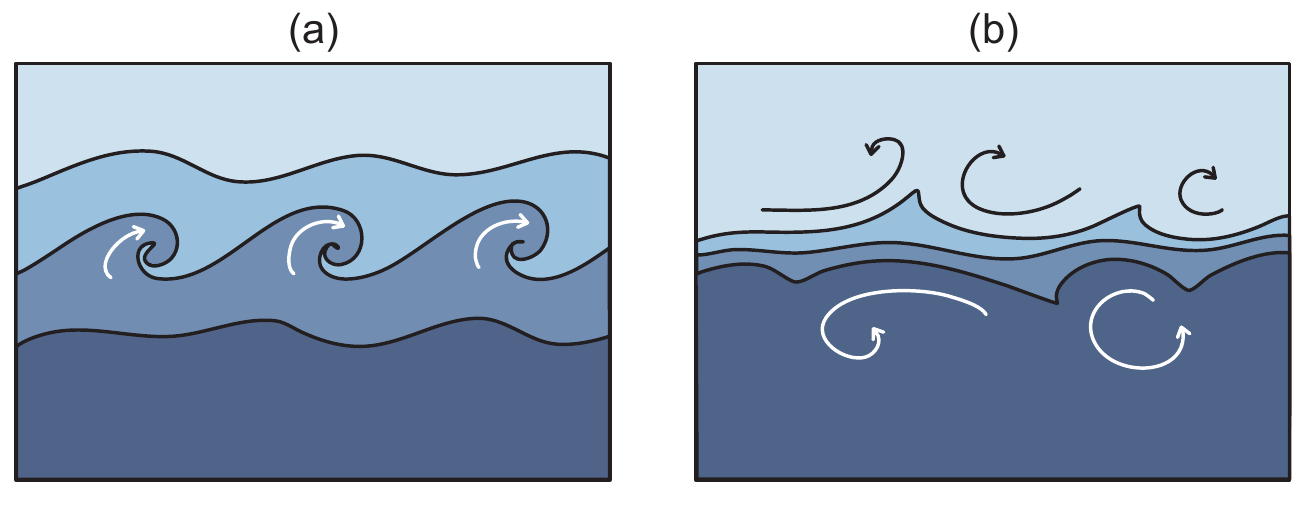}
\caption{Cartoon illustrating overturning (a) and scouring (b) regimes at a density interface \citep{caulfield2021layering}.}
\label{fig:overturn_scour}
\end{figure}

As noted, stratified shelf seas sometimes exhibit a two-layer density structure with relatively homogeneous mixed layers at the top and bottom separated by a stratified thermocline (Figure \ref{fig:seasonal timeseries}). 
There are two paradigms for mixing, at the small scales of density interfaces; `scouring' and `overturning' (see \citet{caulfield2021layering} and references therein). 
The mixing across the density interfaces and the ultimate fate of the density interface depends strongly on which regime the turbulence is in.
The mixing regimes are controlled by the relative strength of stratification and turbulence. This can be quantified by the kinetic energy associated with three-dimensional turbulent eddies and the potential energy associated with the density interface. 
When the stratification at the density interface is sufficiently strong, turbulent eddies do not have kinetic energy to overturn the density interface. 
Instead, if there is a source of turbulence in the surrounding mixed layers, turbulent eddies will `scour' the density interface, pulling characteristic wisps of fluid into the mixed layers (Figure \ref{fig:overturn_scour}b). 
On the other hand, when the kinetic energy associated with the turbulent eddies is large enough, turbulence is able to `overturn' the density interface (Figure \ref{fig:overturn_scour}a). 
\par
The fate of the density interface is intimately tied to the mixing regime; scouring tends to sharpen density interfaces, while overturning tends to mix the interface into a more diffuse state. 
The mixing regime can be quantified by considering the budget for the buoyancy frequency in sorted density coordinates \citep{zhou2017,taylorzhou}: 
\begin{equation}
\frac{\partial N_*^2}{\partial t}=\underbrace{\frac{\partial^2 \kappa_e}{\partial z_*^2}N_*^2}_{A} + \underbrace{2 \frac{\partial \kappa_e}{\partial z_*}\frac{\partial N_*^2}{\partial z_*}}_{B}+\underbrace{\kappa_e \frac{\partial^2N_*^2}{\partial z_*^2}}_{C}.
\label{eq:nstar}
\end{equation}
Here $z_*$ is the height of a vertically sorted isopycnal, adiabatically sorted to monotonically decrease with depth; $N_*^2$ is the buoyancy frequency, Eq.\ \eqref{eq:N}, in sorted height coordinates; and $\rho$ is the potential density, where $\rho_0$ is a reference density. 
The `effective' diffusivity is \citep{winters1996, nakamura1996}
\begin{equation}
\kappa_e=\kappa \left(\frac{\partial z_*}{\partial \rho}\right)^2\left<|\nabla \rho|^2\right>_{z_*},
\end{equation}
where $\kappa$ is the molecular diffusivity and $\left<\cdot\right>_{z_*}$ denotes a spatial average for a fixed $z_*$. Note that this formulation assumes that density is controlled by one variable (temperature or salinity). 
The key advantage of using sorted height coordinates is that $\kappa_e$ is strictly positive. 
A purely laminar flow has $\kappa_e=\kappa$, and turbulence will lead to $\kappa_e>\kappa$ by distorting the isopycnals, increasing the density surface and thereby increasing the density gradients. 
\par
The relative sizes of the terms on the right hand side of Eq.\ \eqref{eq:nstar} can be used to diagnose the mixing regimes. 
Terms $B$ and $C$ in Eq.\ \eqref{eq:nstar} represent translation and diffusion of the sorted density profile, respectively, and since $\kappa_e>0$ this diffusion acts to spread out density interfaces. 
Term $A$ in Eq.\ \eqref{eq:nstar} can be positive or negative, and this term dictates whether turbulence at a density interface is in the scouring or overturning regimes. 
For example, consider a flow in the overturning regime where shear-driven turbulence is generated at a density interface (Figure \ref{fig:overturn_scour}a). 
If the flow above and below the interface is relatively quiescent, $\kappa_e$ could exhibit a local maximum at the density interface with $\partial^2\kappa_e/\partial z_*^2 < 0$.
In this case term $A$ will act to reduce the stratification at the interface. On the other hand, in the scouring regime (Figure \ref{fig:overturn_scour}b) strong stratification will suppress mixing at the interface where $\kappa_e$ will be relatively small.
In this case strong mixing on either side of the density interface can result in a flow with $\partial^2 \kappa_e/\partial z_*^2>0$, in which case term $A$ will act to increase $N_*^2$ and \textit{sharpen} the interface. 
\par
The scouring regime requires a source of turbulence external to the density interface. In natural stratified shelf seas, turbulence generated by bottom friction and surface forcing (e.g.\ wind, waves, and/or convection) generate turbulence below and above mid-water density interfaces. 
Turbulence associated with flow past OWTs could also play this role.
Indeed, as discussed in Section \ref{sec:strat}, horizontal shear can be very effective at generating and maintaining layers in stratified flows.
On the other hand, energetic turbulence driven by the flow past OWTs could be strong enough to overturn and mix natural density interfaces. \par
Thus, it is not clear whether OWTs would generate turbulence in the overturning or scouring regime. 
The distinction will likely depend on the strength of stratification, the speed of flow through offshore wind farms, and the geometry of wind turbines.
As will be expanded on in the following section more research is needed to understand mixing generated by offshore wind infrastructure and their impact from the scale of OWTs to the shelf sea scale.

\section{Flow structure interactions}
\label{sec:infraunstrat}
The dynamics of flows past a range of structures has seen significant research over the last 50 years, both due to immediate real-world applications and the increase in computational resources and experimental measurement fidelity \citep{williamson1996vortex}.
Here work studying the dynamics of unstratified and stratified flow past infrastructure, relevant to the offshore wind sector, are integrated and reviewed.
\newline
\subsection{Unstratified Flow}
\begin{figure}
\centering
\includegraphics[width=\textwidth]{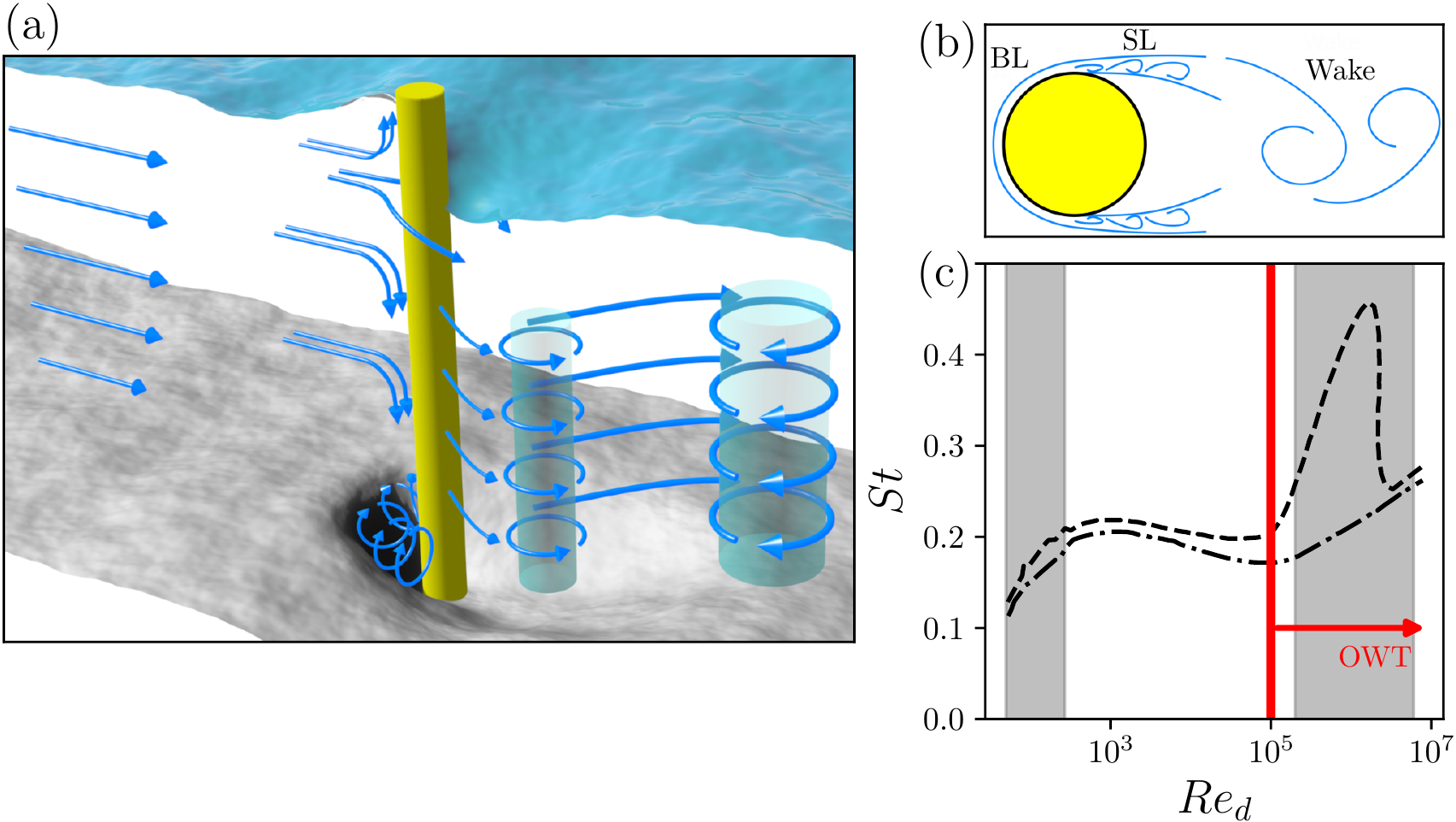}
\caption{Typical flow past a cylinder.
Panel (a) shows the finite-depth 3D flow past a cylinder over an erodible bed, depicting the shedding Karman vortices (KV). 
Note, although not discussed herein, the run-up/depression at the water surface upstream/downstream of the cylinder and formation of horseshoe vortices cause scour at the bed, which may destabilise structures \citep{matutano2013scour}.
Panel (b) shows the three shear layers comprising the cylinder flow: BL denotes the boundary layer, and SL denotes the shear layer.
Panel (c) shows the typical $Re_d-St$ relationship for flow past a cylinder.
Shaded regions represent the 2D laminar vortex shedding regime ($49 \lesssim Re_d \lesssim 192$) and the trans-critical flow regime ($2\times10^5 \lesssim Re_d \lesssim 6\times10^6$).
Upper curve represents flow for a smooth cylinder, lower curve for a rough cylinder. 
Curves are based upon the experiments reviewed by \citet{lienhard1966synopsis}.
The red line indicates the expected minimum Reynolds number of an OWT.
}
\label{figure:unstratifiedCylinderFlow}
\end{figure}

\begin{figure}
\centering
\includegraphics[width=\textwidth]{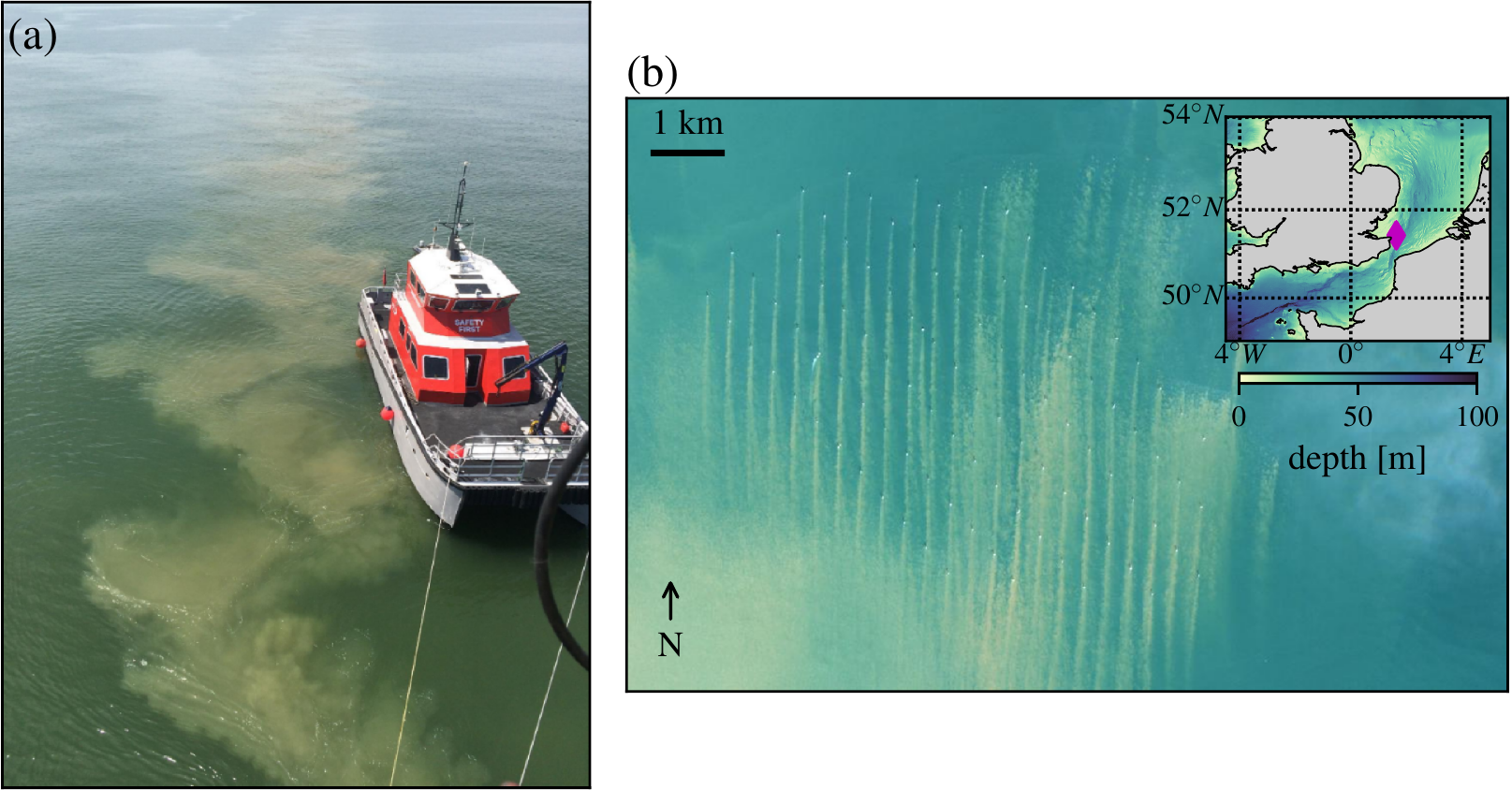}
\caption{
Sediment plumes generated by cylindrical structures. a) Wake and turbid plume showing the coherent Karman street from the lee of a metocean mast, image courtesy D. van der Zande.
b) 
Satellite imaging of turbid plumes show they extend over multiple kilometers, individual monopile diameters are $d=\SI{4.3}{m}$, image courtesy R. Forster \citep{rodney18plume}. 
With average monopile spacing of 1 km, interactions of the turbid plumes are clearly visible.
}
\label{fig:plumes}
\end{figure}

Flow past cylindrical structures are well studied, owing to their geometrical simplicity and vast engineering importance; such studies are directly analogous to flow past OWT foundations, such as monopiles.
Unsteady vortices shed by cylinders can lead to vibration, acoustic noise, resonance, and ultimately structural failure.
Shed vortices form coherent wakes that are spatially vast, and can be detected several hundred diameters downstream of their source, depending on the background flow conditions (e.g. turbulence properties).
Cylinder wakes fundamentally alter flow conditions as a source of anthropogenic turbulence (mixing), particularly coherent along the cylinder axis.
Dynamics are complex due to the interaction of at least three shear layers; the boundary layer, shear layer, and wake (Figure \ref{figure:unstratifiedCylinderFlow}).
In unstratified waters, flow structure is dependent on the cylinder roughness, end conditions, freestream conditions, and the ratio of inertial to viscous forces, the cylinder Reynolds number $Re_d$,
\begin{equation}\label{eq:red}
    Re_d = \frac{u_\infty d}{\nu},
\end{equation}
where $u_\infty$ denotes the freestream velocity, $d$ the cylinder diameter and $\nu$ the fluid viscosity.
The dependence of the flow structure on $Re_d$ arises due to transitions in the different shear layers.
The most well known instability arising from the flow past a cylinder is the Karman vortex (KV) \citep{williamson1996vortex}, which develops for $Re_d \gtrsim 49$ (below which the flow is steady and laminar).
The Karman instability is associated with alternating 2D vortices shed from either side of the cylinder, aligned with the cylinder axis, and is a consistent feature of even high Reynolds number flows.
The unsteady KV is characterised by the dimensionless frequency, the Strouhal number $St = f_\text{KV} d/u_\infty$, where $f_\text{KV}$ is the frequency of vortex shedding. $St$ varies with $Re_d$ depending on shear layer transitions, as shown in Figure \ref{figure:unstratifiedCylinderFlow}c. The KV is a dominant feature of flow past a cylinder (Figure \ref{fig:plumes}a). 
\par
The typical $Re_d$ of shelf sea currents past offshore wind foundations (Figure \ref{fig:structures}) 
is at least $Re_d \gtrsim 10^5$, estimated assuming a small \SI{5}{m} diameter monopile with a minimum tidal velocities of \SI{0.02}{\m \per \second} \citep{vindenes2018analysis}.
This minimum Reynolds number is sub-critical (see Figure \ref{figure:unstratifiedCylinderFlow}), indicating that shear layer instabilities may be present, manifesting as Kelvin-Helmholtz (KH) type instabilities. 
Three dimensionality is a dominant feature of cylinder wakes at these high Reynolds numbers.
Critical transition occurs between $2\times 10^5 < Re_d < 5\times 10^5$, and is associated with boundary layer transition to turbulence which causes the separation point to occur further downstream on the cylinder surface.
For a smooth cylinder, asymmetric separation-reattachment of the boundary layer either side of the cylinder causes a sudden increase in vortex shedding frequency (Figure \ref{figure:unstratifiedCylinderFlow}).
Super-critical flow is associated with $Re_d > 5\times 10^5$, where symmetric separation bubbles and turbulent boundary layers are present on both sides of the cylinder.
Roughness effects are felt primarily in the critical transition regime; surface roughness causes earlier transition to turbulence and bypasses the asymmetric regime (Figure \ref{figure:unstratifiedCylinderFlow}b).
\par
With distance $x$ downstream, the far-wake $x/d \gtrsim 50$ is particularly sensitive to freestream conditions. 
The spectral energy of the KV (energy associated with $f_\text{KV}$) decays downstream of the cylinder and the wake width grows like $\sqrt{x}$ (under low levels of freestream turbulence) and is approximately self-similar \citep{ghosal1997numerical}.
A secondary vortex street emerges at a lower frequency than the KV, arising due to the merging of vortex pairs, or via hydrodynamic instability of the mean flow, depending on the flow Reynolds number \citep{jiang2021formation}.
Coherent structures in the far-wake can also arise due to non-linear interactions between freestream structures and the KV. 
\citet{cimbala1990effect} and \citet{williamson1993new} found that the interaction between the KV and freestream waves could lead to resonant peaks in spectral energy associated to their non-linear interaction, indicative of hydrodynamic instability.
These peaks can be detected far downstream of the cylinder, $x/d > 300$, indicating that non-linear dynamics have a large effect on the wake even far downstream.
It is therefore unsurprising to observe non-linear interaction between wake effects from multiple monopiles in offshore wind farms (Figure \ref{fig:plumes}).
Under high freestream turbulence the wake spreads more rapidly, depending on the background turbulent intensity and the turbulent integral length scale. 
\citet{eames2011growth} demonstrated that the growth rate of the wake increases from $\sim \sqrt{x}$ to $\sim x$ when the wake deficit velocity is approximately equal to the background turbulent intensity, and the integral length scale of the turbulence is comparable to the cylinder diameter.
When subject to high turbulence a cylinder wake will dissipate downstream more rapidly and diffuse into the background turbulence. 
In contrast, when background turbulence levels are low the turbulence generated in the wake of a cylinder can persist hundreds of diameters downstream.
\par
In current shallow-water offshore wind farms, where levels of turbulence are high, wakes have been observed at least \SI{1}{km} in length (e.g. Figure \ref{fig:plumes}).
The formation of turbid plumes, where suspended sediment is trapped and transported in KVs and secondary vortex streets, is correlated with high levels of TKE \citep{grashorn2016karman}.
As the wakes pass through the wind farm plumes are observed to spread and interact (Figure \ref{fig:plumes}b).
Wake spreading indicates that the effects of monopile wakes are not limited to a short narrow region downstream.
Although KV are expected to decay for $x/d \approx 60$ \citep{jiang2019transition}, observed plume interaction at large lengths, $x/d > 150$, suggests wake-wake interactions are important.
However, conditions vary considerably between the current shallow coastal offshore wind farms and the deep water future development sites.
It is to be expected that in deep water sites, with lower background turbulence, that wakes may be even larger.
\newline
\subsection{Stratified Flow}\label{sec:strat}
Relatively few studies have investigated stratified flow interaction with vertically oriented cylinders, analogous to proposed offshore wind foundation deployment in seasonally-stratified shelf seas. 
This may be in part due to similarity between unstratified and stratified flow past vertical cylinders at very low Reynolds, $Re_d\lesssim 45$ \citep{meunier2012stratified}, where wakes are inherently two-dimensional.
However, differences arise when three-dimensionality is present in the wake, which is certainly characteristic of the high Reynolds number flows associated with offshore wind farm infrastructure.
Three-dimensionality in the wake of vertical cylinders is important as it can lead to fundamental reorganisation of stratified fluids. 
Layering can emerge where such a flow can lead to multiple intermittent regions of fairly constant density neighbouring thin interfaces with steep density gradients \citep{bosco2014three}.
This layering behaviour may arise from any process that produces spatially periodic mixing in the vertical direction \citep{thorpe2016layers}. 
\begin{figure}
\centering
$\vcenter{\hbox{
\begin{tikzonimage}[width=0.6\textwidth]{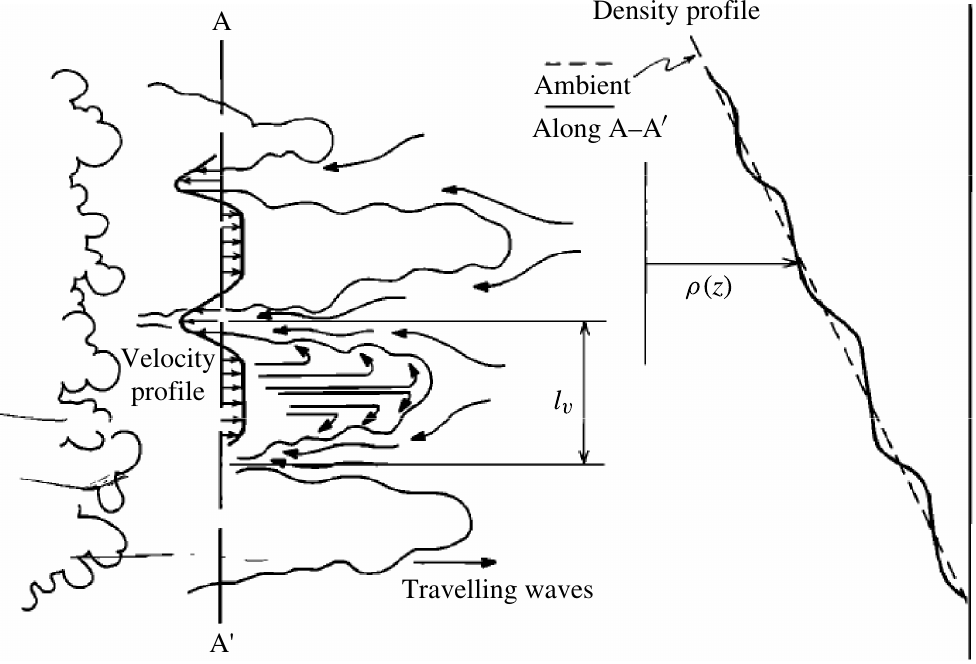}
\node at (0.05, 0.965) {\large(a)};
\end{tikzonimage}
}}$
\hfill
$\vcenter{\hbox{
\begin{tikzonimage}[width=0.35\textwidth]{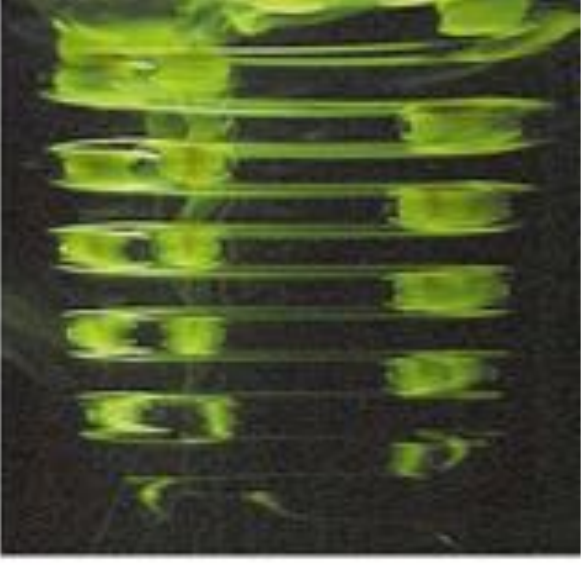}
\node at (0.05, 1.05) {\large(b)};
\end{tikzonimage}
}}$
\caption{Layering in a linearly stratified fluid emerging from a turbulent cloud, with the resulting density profile across the intrusions (a), from \citep{maffioli2014evolution}.
A Dye visualisation of the zig-zag instability is presented in panel (b), from the experiments of \citet{billant2000experimental}.
}
\label{fig:layering}
\end{figure}
\par
Cylinder wakes are prone to layering due to their increase in both horizontal shear, normal to the direction of stratification, and organised vertical vorticity, parallel to the direction of stratification, through the formation of KV in particular.
The susceptibility of a stratified flow, perturbed by moving cylinders, to layering is dependent on the mechanisms that generate layers, which vary considerably depending on the type of forcing \citep[e.g. oscillatory versus continuous stirring, ][]{thorpe2016layers}. 
Experiments have demonstrated that layers can develop from the `cloud' of stratified turbulence that results from dragging vertical cylinders through a stratified flow, equivalent to flow through a wind farm (Figure \ref{fig:layering}).
In the experiments of \citet{maffioli2014evolution} intrusions developed from local regions of mixed fluid, which grew until their length scales were approximately balanced by $l \sim \tilde{u}/N$, where $\tilde{u}$ represents a characteristic velocity scale of the turbulence.
The buoyancy frequency may also be used to define a Froude number, the ratio of inertial to gravitational forcing, where $L$ is some length scale over which flow velocity and $N$ are averaged,
\begin{equation}
    Fr = \frac{\tilde{u}}{NL}.
\end{equation}
Thus, in the experiments of \cite{maffioli2014evolution}, packets of turbulent fluid grew until they collapsed under gravity ($Fr \sim 1$) and spread outwards as pancakes, triggering horizontally propagating internal waves.
The outward spread of intrusions generates a layered density profile, where well-mixed intrusions are neighboured by thin regions of strong density gradient.
It is important to note here that the cylinders are not critical to the reorganization of stratification; they only act as the source of turbulence, via horizontal shear, in the stratified flow.
Addition of purely horizontal shear to simulations of stratified shear flow can lead to coherent vertical vortices \citep{basak2006dynamics}, analogous to the KV and KH instabilities of a cylinder wake. 
The coherent vertical vortices exhibit pairing, tearing, and amalgamation, resulting in a complex braided vorticity structure that ultimately leads to vertical variability.
Layers can subsequently develop as intrusions, coupled with internal waves \citep{basak2006dynamics}, once again with a characteristic vertical scale of the order of $\tilde{u}/N$, when these two quantities are estimated appropriately.
\par
Indeed, provided the flow Reynolds number is sufficiently large, there is accumulating evidence,  that  horizontal shear (and hence vertical vorticity) in vertically stratified fluids inevitably forms layers on this scale. Dating back to the first theoretical analysis of \cite{billant2000theoretical}, there is clear evidence that such flows are prone to a class of `zig-zag' instabilities. These instabilities imprint the $\tilde{u}/N$ vertical scale on the flow, and can be connected directly
to the inherently nonlinear layered structures that develop at finite amplitude\citep{lucas2017}. Furthermore, several studies\citep{deloncle2008,waite2008,augier2015} have demonstrated that the breakdown of these vertical vortices in a stratified fluid introduce a new, inherently stratified route to turbulence in a stratified fluid, and hence substantially enhanced mixing.
\par
Whilst work on layering has been restricted to small scale experiments and comparatively low Reynolds numbers (at least by oceanographic standards), it has provided oceanographers with vital information on the fine-scale density structure of the ocean. 
However, it is unclear how turbulence generated at high Reynolds number, by flows past offshore wind foundations, will interact with the  essentially two-layer density profile of seasonally stratified shelf seas.
In addition to the high Reynolds number, offshore wind foundations are a similar width to the typical thermocline thickness.
This contrasts with previous work where turbulence was either generated by structures with length scales two orders of magnitude smaller than the density gradient length scale \citep{maffioli2014evolution}, or (as in the `zig-zag' instability studies mentioned above) the unstable horizontal shear/vertical vorticity is embedded initially in a linearly stratified fluid with close to constant buoyancy frequency.
Where KV shed from offshore wind foundations are of the same order diameter as the density gradient length scale, flow mixing and density profile reorganisation may be fundamentally different, and this generic stratified flow geometry with a range of key characteristic length scales is very poorly understood, even in highly idealised circumstances.
\par
Expansion of the offshore wind sector to deeper waters is predicated on development of floating foundations, which are finite in depth and anchored to the seabed.
Finite depth structures induce complex dynamics within stratified fluids, especially if structures intersect sharp density gradients, i.e the thermocline (Figure \ref{fig:structures}).
A good example case study of the impact of finite depth obstacles on stratified flow, is flow past a horizontal cylinder.
At high Froude numbers dynamics are similar to unstratified flow. 
But as the Froude number decreases several new regimes occur \citep{boyer1989linearly}.
Under stratification internal waves can be generated not only by the structure but also the KV.
This in turn can increase drag, and thus the amount of mixing, by up to 100\% \citep{arntsen1996disturbances}.
The drag coefficient of spheres in stratified flows also varies by up to 100\%, when the vortex shedding frequency tends to the buoyancy frequency  \citep{cocetta2021stratified}.
A variety of fixed and floating foundations may be susceptible to such additional drag, where vortices shed are not necessarily aligned with the direction of stratification (e.g. jackets and semi-submersible foundations, Figure \ref{fig:structures}). 
\par
\begin{figure}
    \centering
    \includegraphics[width=0.8\textwidth]{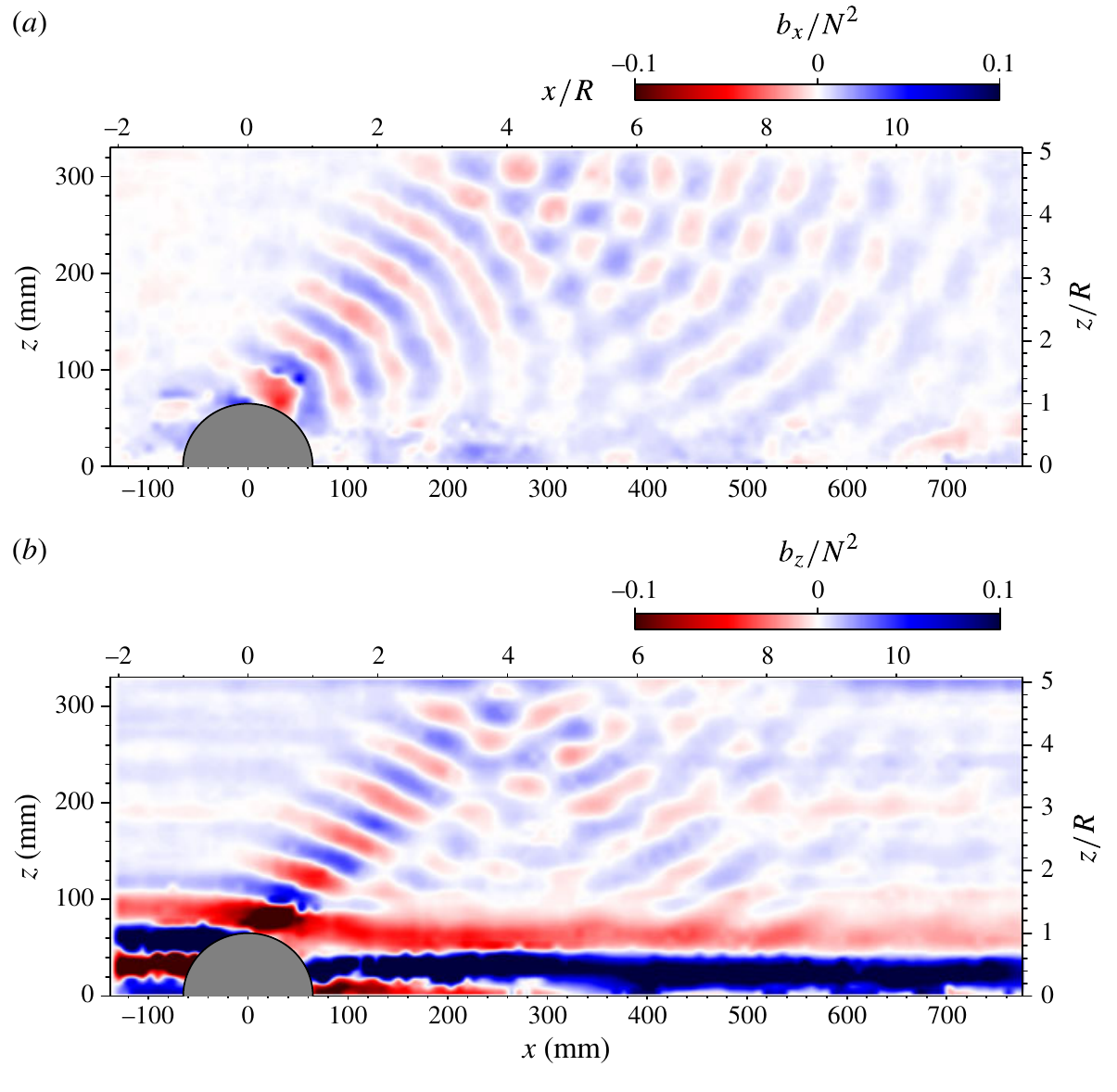}
    \caption{Stratified flow over a hill, taken from the experiments of \citet{dalziel2011structure}. Internal waves can be observed in the lee of the structure propagating vertically and horizontally, clear in both the horizontal buoyancy gradient $b_x = - \partial \rho / \partial x$ and vertical buoyancy gradient $b_z = - \partial \rho / \partial z$. The wake is also particularly apparent in the $b_z$ field, demonstrating the existence of significant mixing). $N$ is the background buoyancy frequency and $R$ denotes the height of the hill.}
    \label{fig:stratified_topography}
\end{figure}

Through considering reflectional symmetry, flow past floating foundations may be analogous to oceanic and atmospheric flow past sea-mounts and hills.
Stratified flow past such obstacles displaces fluid and leads to both vertically and horizontally propagating internal waves, as well as a significant downstream wake (Figure \ref{fig:stratified_topography}).
If an obstacle is wide compared to its height, most fluid impinging on the obstacle passes over it, otherwise flow can pass around it.
For low Froude numbers, where $L$ is scaled with obstacle depth, flow blockage by internal waves creates additional drag, which is a very effective momentum sink on the impinging flow. 
Flow blockage can increase the drag force by 1-2 orders of magnitude compared to unstratified flow over the same obstacle \citep{smith1978measurement,castro1990obstacle,cummins1994simulated}.
In all cases understanding drag increase  is critical to evaluate both the hydrodynamic loads placed on foundations and anchors, and the mixing in stratified shelf seas, associated with the flow past offshore wind foundations.
Such wave generation and propagation also suggests the possibility that the effect of flow structures can be both local and non-local, as the emitted waves may transport momentum flux significant distances until they `break'.
\newline
\subsection{Mixing of Stratified Shelf Seas by Offshore Wind Foundations}
\label{sec:inframix}
\begin{figure}
\centering
\includegraphics[width=0.9\textwidth]{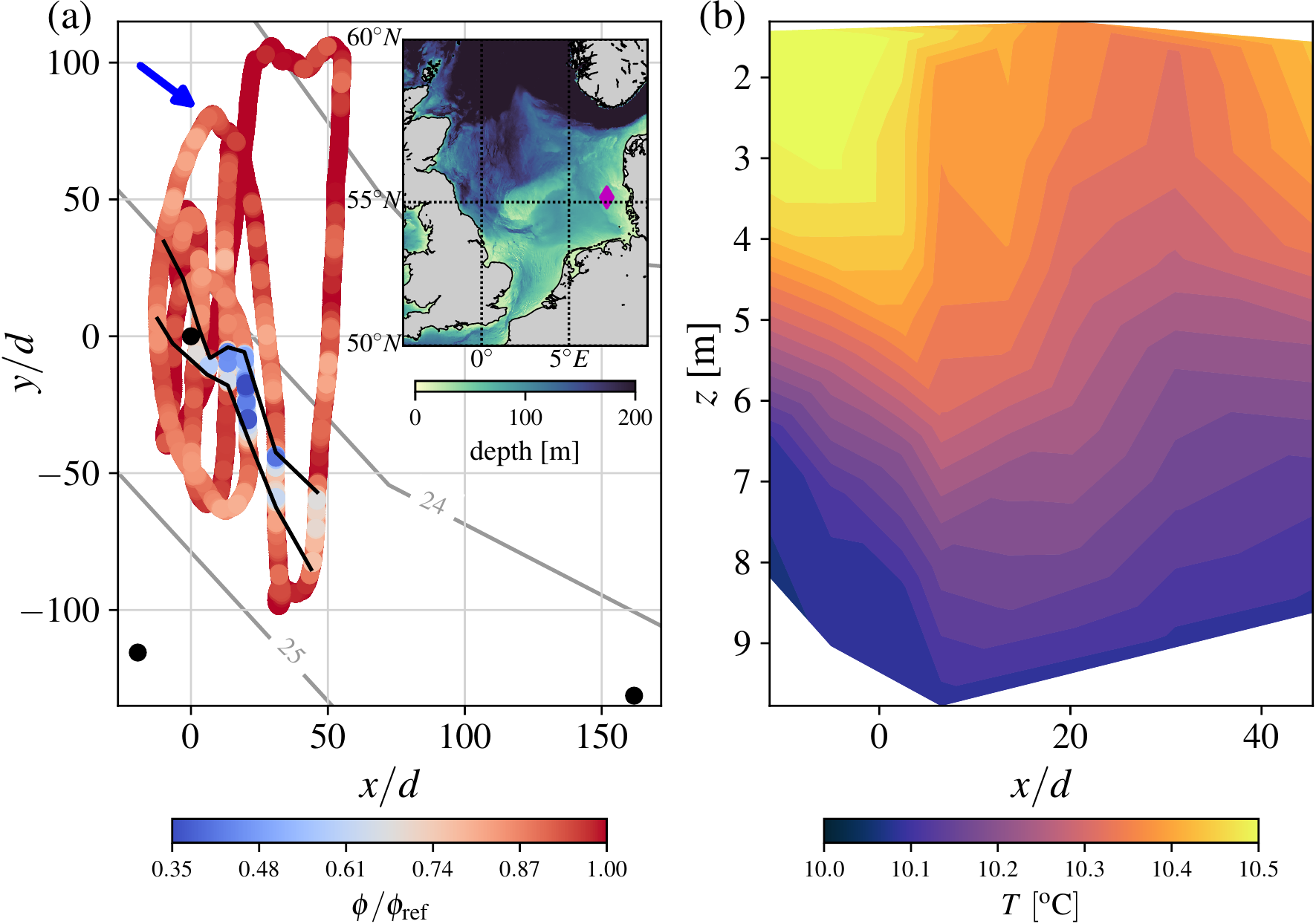}
\caption{Observations of the turbulent stratified wake downstream of a monopile, derived from the survey conducted by \citep{schultze2020increased}.
The monopile of diameter 6 m, is located at 55.16$^\circ$N 7.07$^\circ$E on the leading edge, relative to local flow, of the DanTysk wind farm. The wind farm is marked by the diamond in the inset figure of (a).
(a) depicts the potential energy anomaly, $\phi$ around the monopile.
$\phi$ is normalised by its background reference value, $\phi_\text{ref}$.
The approximate flow direction is indicated by the blue arrow, and neighbouring monopiles are marked by black circles. 
The wake is clearly visible by the blue regions, and approximately enclosed by the black lines.
(b) shows the spatial development of the temperature field in the flow before and the wake after the monopile.
Evolution of the temperature field is obtained by averaging data in the regions enclosed by the black lines in (a).
Data are from \citet{schultze_data}.
}
\label{fig:schultze_data}
\end{figure}
To date there has been only two limited studies observing offshore wind foundation induced mixing of stratified waters \citep{floeter2017pelagic,schultze2020increased}.  
These studies have been restricted to developments in aperiodically stratified regions of freshwater influence, in shallow water (depths of approximately \SI{40}{m} and \SI{24}{m}, respectively).
Neither studies have investigated the potential impacts on seasonally, or permanently, stratified shelf seas.
\par
\citet{floeter2017pelagic} performed surveys on two wind farms in the German Bight, North Sea.
Water property transects through the wind farm revealed a consistent weakening of stratification near the centre.
Effects extended into the surrounding area by approximately half the diameter of an ambient tidal excursion.
However, it was unclear how much of this was due to `infrastructure' turbulence from turbine foundations rather than natural topological effects.
In addition to reduced stratification, \citet{floeter2017pelagic} also measured local upwelling at the edges of the wind farms, similar to those observed near islands in stratified waters \citep[e.g.][]{simpson1982mixing}.
Shallow water model studies of the effects of offshore wind farms on local oceanic circulation patterns have shown arrays of infrastructure induce strong horizontal shear in wind stress leading to local regions of upwelling and downwelling \citep{brostrom2008influence,paskyabi2012upper}, consistent with the observations of \citet{floeter2017pelagic}. 
\par
Field measurements were also taken by \citet{schultze2020increased} who measured the stratified wake from an offshore monopile at the leading edge, with respect to local flow, of the DanTysk wind farm (Figure, \ref{fig:schultze_data}).
The monopile wake spread to a width 10 times the (\SI{6}{m}) diameter of the monopole and had reduced the potential energy anomaly by up to \SI{65}{\%} at a downstream distance of $x/d \approx 20$ (Figure \ref{fig:schultze_data}).
The full distance required to return to pre-monopile conditions was not captured by the survey, even after approximately \SI{300}{m}, a distance over 50 times the \SI{6}{m} monopile diameter.
The survey, clearly demonstrates that turbulence generated by monopiles reduces stratification.
Complementing the survey, \citet{schultze2020increased} used Large Eddy Simulations (LES) to model flow past a single monopile which was simulated under different levels of background stratification.
TKE dissipation rate was found to be up to two orders of magnitude larger in the thermocline than when monopiles were not present.
High TKE dissipation rate persisted far downstream of the cylinder and, at $x/d>40$, was still an order of magnitude larger than without monopiles at the thermocline depth. 
Further, the TKE dissipation rate was greater than that generated by the bottom boundary layer.
\par
During a period of stronger stratification, \citet{schultze2020increased} conducted a second survey at the opposite end of the DanTysk wind farm.
The second survey was less conclusive than their previous measurements, finding that no clear signal from the wake could be separated from background variability. 
This may be because stratification was strong enough to suppress the growth and interactions of the wakes, although \citet{schultze2020increased} only sampled the wake at distances greater than \SI{200}{m} (or 33.3 diameters) from the monopile.
It is therefore unclear if this is a result consistent with the earlier survey in Figure \ref{fig:schultze_data}.
\par
\begin{figure}
    \centering
    \includegraphics[width=\textwidth]{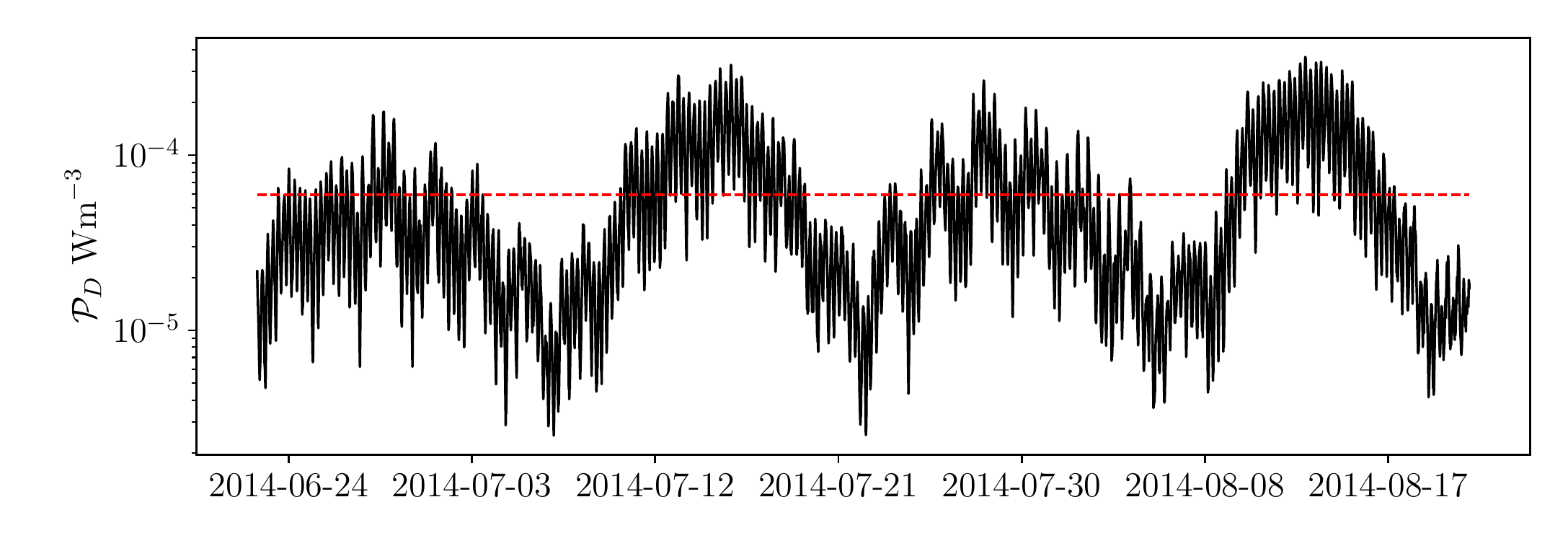}
    \caption{Mean TKE production by a $d=10$ m offshore monopile.
    TKE production is based on direct velocity measurements averaged over the thermocline, \SIrange{20}{50}{m} in the Celtic Sea between 22/06/2014 and 20/08/2014, Figure \ref{fig:seasonal timeseries} \citep{Scannellpreprint}. 
    Turbulence production Eq.\ \eqref{eq:power} is closed by assuming $c_D = 1$, $\rho_0=$\SI{1000}{\kg m^{-3}} and $D=$\SI{200}{m}.
    The power averaged over the full time period is marked in red.}
    \label{fig:powerlosses}
\end{figure}
Leading order arguments for the mixing induced by offshore wind foundations have been explored by \citet{rennau2012effect} and \citet{carpenter2016potential}, here they are reviewed.
Models start by assuming that the turbulence produced by foundations is equal to the power lost to drag. 
Over an arbitrary vertical layer of fluid, $L$, the power lost to drag is given by
\begin{equation}
    P_D = \frac{1}{2} \rho c_D d L \langle|\mathbf{u}|^3\rangle_L,
\end{equation}
where $c_D$ denotes a drag coefficient and $\langle|\mathbf{u}|^3\rangle_L$ the velocity magnitude cubed, averaged over the layer $L$. 
Whilst \citet{rennau2012effect} and \citet{carpenter2016potential} assumed that $c_D$ is a constant in reality it is variable and dependent on the Reynolds number, surface roughness and structure geometry. 
The drag coefficient is also likely a function of depth and time and will vary with background shear. 
Crucially $c_D$ is also dependent on the Froude number, especially for highly stratified flows (see Section \ref{sec:strat}).
At present there is no parametrisation of the Froude number dependence of the drag coefficient for vertical infrastructure in stratified flows. We therefore assume $c_D = 1$ as per the ``high drag'' case of \citet{carpenter2016potential}, although this estimate does not account for the potential effects of the Froude number.
\par
A wind farm comprises many turbines, potentially supported by a range of foundations.
Turbines are separated by a distance of approximately $10D$ \citep{howland2019wind}, where $D$ is the rotor diameter.
The average production of TKE per unit volume $\mathcal{P}_D$ over a layer of thickness $L$ is therefore:
\begin{equation}
\label{eq:power}
    \mathcal{P}_D = \frac{1}{2} \frac{\rho_0 c_D d \langle|\mathbf{u}|^3\rangle_L}{100 D^2}.
\end{equation}
Here it is assumed that the power lost to drag at a turbine foundation acts to increase TKE production over the full area occupied by a foundation. 
In reality this production will be localised to the comparatively narrow wake, and will vary by several orders of magnitude in space.
\par
Here the potential production of TKE $\mathcal{P}_D$ per unit volume, generated by the addition of a 10 m diameter monopile, is estimated using the natural conditions, averaged over the thermocline thickness, at the CaNDYFLOSS experiment site (Figure \ref{fig:seasonal timeseries}).
Power generated by flow past such a potential foundation is strongly dependent on the flow velocity, with low frequency oscillations associated with neap-spring tide cycles and higher frequencies associated to daily tidal cycles (Figure \ref{fig:powerlosses}). 
The potential power added to the thermocline, $\mathcal{P}_D$, varies between approximately \SIrange{2e-6}{4e-4}{\watt \per \m \cubed} with an average value of \SI{5.9e-5}{\watt \per \m \cubed}.
Over the same period, at the same site, the natural background dissipation rate ($\varepsilon_B$) at the thermocline is \SI{3.5e-5}{\watt \per \m \cubed} \citep{Scannellpreprint}.
In this approximate calculation, the additional (average) turbulence production is therefore 69\% greater than background dissipation without monopiles present: $\mathcal{P}_D \approx 1.69 \varepsilon_B$.
\par
Under several very strong assumptions, the relation between TKE produced by offshore wind foundations and how it is balanced by viscous dissipation and work performed on the buoyancy field may be expressed, from Eq.\ \eqref{eq:prod}, by 
\begin{equation}
\label{eq:prod_osw}
    \mathcal{P} = \mathcal{P}_D + \mathcal{P}_B = (1 + \Gamma)\varepsilon_D + (1 + \Gamma)\varepsilon_B,
\end{equation}
where subscript $D$ represents contributions from structures (drag) and subscript $B$ represents all other (background) contributions to TKE production ($\mathcal{P}$) and dissipation rate ($\varepsilon$). \par
For simplicity, the  turbulent flux coefficient can be assumed equal for each component of $B$ and $\varepsilon$ such that $\Gamma = B/\varepsilon = B_B/\varepsilon_B = B_D/\varepsilon_D$. 
In deriving \eqref{eq:prod_osw} it is also assumed that OSW infrastructure contributes linearly to $\mathcal{P}$, and is simply related to the structure-induced TKE dissipation rate by $\mathcal{P}_D = (1+\Gamma)\varepsilon_D$. 
In the present example we have shown that $\mathcal{P}_D \approx 1.69 \varepsilon_B$, such that
\begin{equation}
    \mathcal{P}_D = (1 + \Gamma) \varepsilon_D \approx 1.69 \varepsilon_B.
\end{equation}
Under the simplest conventional assumption that $\Gamma = 0.2$, as typical at the pycnocline in seasonally stratified waters (see Section 2), we obtain $\varepsilon_D \approx 1.4 \varepsilon_B$, such that the total TKE dissipation rate, $\varepsilon = \varepsilon_B + \varepsilon_D$, is at least \SI{140}{\%} higher at the thermocline when OSW infrastructure is present. As noted, several sweeping assumptions have been made to arrive at this estimate, which inevitably has a large amount of implicit uncertainty. It is reasonable to suppose that this estimate is likely to be a conservative lower bound on the effect of OSW infrastructure on turbulent dissipation and mixing at the thermocline for at least two reasons. The first is  that there is clear potential for a strongly nonlinear effect of such infrastructure on the dissipation rate. Secondly,  setting $\Gamma \simeq 0.2$ may well be an under-estimate of the vigorous overturning mixing likely to be triggered in the wake of 
such infrastructure \citep{caulfield2021layering}.
Nevertheless, these simple conservative estimates show that foundations produce turbulence at levels that will clearly affect the leading order balance of TKE transport, even when normalised by the \emph{total} area between offshore wind foundations.
Thus, offshore wind has the potential to  impact directly the stability of seasonally-stratified shelf seas (Figure \ref{fig:marginal}).
\par
The dissipation rate estimate indicates that OSW infrastructure is likely to affect the leading order balance of TKE transport in vicinity of the windfarm, but the implications of this on regional-scale fluid dynamics is more challenging to evaluate. 
To address this, \citet{carpenter2016potential} derived similar models for TKE production as \eqref{eq:power}, and constructed arguments based on estimates of the different timescales in the flow. In particular, timescales of OSW-induced mixing, and timescales of flow parcels convecting through the windfarm. 
By comparing different estimates of these timescales \citet{carpenter2016potential} theorised the extent that the water column would mix as a parcel passed through the array. 
It was concluded that small developments in shallow waters (Bard 1 and Global Tech 1, German Bight) were unlikely to affect stratification, where TKE production from the bed and free surface dominate.
However, despite the simplification of the models and assumptions made, large scale developments in deep water were recognised to have potential for significant impact \citep{carpenter2016potential}.
\par
\citet{rennau2012effect} adopted regional-scale numerical models to assess the effects of offshore wind development in the Baltic Sea, where tidal currents are minimal and stratification are driven by dense saline currents beneath fresh water.
They modified a two-equation (RANS) turbulence closure scheme to capture the enhanced mixing arising from OSW installations and found that current installations did not have a significant effect on regional stratification (though cautioned that future development, covering more of the Western Baltic Sea, could lead to significant impacts, such as reduced bottom salinity). However these closure schemes are known to be overly dissipative in stratified flows \citep{hewitt2005prediction}, and demonstrated to generate overly diffuse pycnocline structures in realistic simulations \citep{Luneva19}. These schemes are therefore improperly conditioned, at present, to address the impact of OSW infrastructure in stratified seas. 

\section{Discussion}
The scale of planned offshore wind energy industry is much greater than past, and existing, sea use. 
Installed offshore wind capacity will increase by 600\% in the next decade (Figure \ref{fig:globalwind}a), requiring an extra $\sim 20,000$ 10 MW+ turbines.
Fixed, and floating, offshore wind infrastructure will penetrate the thermocline, adding `anthropogenic' mixing on top of natural mixing.
The impact of such infrastructure will be fundamentally different from existing sea use.
For example, even large surface vessels have comparatively small drafts, $\sim\SI{10}{m}$ \citep{golbraikh2020model,nylund2020situ}.
Further, whilst offshore oil and gas platforms are similar to offshore wind infrastructure, spanning the thermocline, only $\sim6,500$ platforms have been installed, globally, over the last 75 years \citep{schneider2010foundation}. 
Therefore, planned offshore wind developments will add new and large scale infrastructure sources of turbulent mixing in seasonally stratified seas.
This discussion reviews our current knowledge gaps, frames potential impacts on shelf sea dynamics and thus marine ecosystem functioning and highlights routes for sustainable growth of the offshore wind sector.
\newline
\subsection{Infrastructure Mixing}
The mixing of stratified waters by offshore wind infrastructure is poorly understood.
As evidenced in Section \ref{sec:inframix} there is a dearth of research on high Reynolds number stratified flow past vertical structures, which is vital for understanding and parameterising `infrastructure' mixing processes in natural environments.
The problem is particularly complex due to its scale; laboratory experiments and fully resolved numerical simulations are limited to relatively low Reynolds number flows, certainly by comparison to the real oceanographic flows.
In all scenarios of fixed or floating OWTs, fine scale vertical density structure and enhanced mixing is anticipated due to horizontal shear generated by the flow past the infrastructure. The $\mathcal{O}\left(10^{-2}-10^0\right)$ m length scales associated with the resulting turbulence will be unresolved by low resolution numerical simulations.
Wakes from individual structures may persist for $\mathcal{O}\left(10^2-10^3\right)$ m downstream.
From individual wind turbines to a single offshore wind farm,  infrastructure adds a wide range of length scales of $\mathcal{O}\left(10^1-10^4\right)$ m.
Further, multiple multiple wind farms are distributed at shelf-wide scales.
The vast range of length scales present in flow past offshore wind farms necessitates a variety of modelling techniques.
\par
Although coarse numerics, i.e. Large Eddy Simulation (LES) or other more sweeping turbulence closure schemes, may be desirable modelling strategies for high Reynolds number flow \citep{rennau2012effect,carpenter2016potential}, without a firm understanding of the underlying physics, and robust datasets for validation, low resolution numerical models should be adopted with care, particularly for stratified (and therefore highly anisotropic) flows \citep{hewitt2005prediction,khani2015large}. 
Integration of modelling disciplines ranging from Direct Numerical Simulation (DNS) of idealised flows to regional scale ocean circulation models is therefore required to determine the impact of offshore wind farms on seasonally stratified shelf seas. 
Field surveys are essential to support and validate physical and numerical studies.
However, the limited work to date \citep{floeter2017pelagic,schultze2020increased} has suffered from uncertainties, where either effects of infrastructure were difficult to discern from topographical effects or where wakes were difficult to separate from background variability. 
While both studies concluded effects of infrastructure on mixing may be large, further surveys are required to support this. 
In addition there is a clear need for repeat and `before and after' surveys, as noted by \citet{van2020effects}.
\par
There are also key research questions regarding the geometry of offshore wind infrastructure, and how this impacts mixing. 
OWTs shed vortices with horizontal scales that are comparable to the thickness of the thermocline.
Such mixing contrasts with previous work studying mixing by thin vertical cylinders, several orders of magnitude smaller in diameter compared to density length scales. 
It is vital to understand how large scale vortex structures interact with a relatively thin thermocline; mixing processes may well be fundamentally different to those studied before, with complex spatio-temporally variable vortex, turbulence and mixing dynamics.
In addition, future floating technology raises further questions regarding the impact of geometry on mixing. 
It is expected that spar-buoy designs (Figure \ref{fig:structures}) will act in a similar way to monopiles, given they penetrate through the thermocline and into the well-mixed deep water.
However, semi-submersible (or any other small-draft designs) will introduce non-trivial effects by intersecting the thermocline. 
Research of stratified flow over finite topography, similar to small draft floating OWT, has demonstrated that baroclinic effects can enhance drag by up to two orders of magnitude (Section \ref{sec:strat}). 
Here, floating structures will introduce infrastructure mixing via shed lee waves, internal waves, blockage effects, and wake-wake interactions in the case of semi-submersible designs.
\par
Crucially, a lack of insight into key multi-scale mixing processes adds uncertainty to current attempts to quantify the impact of infrastructure mixing on shelf sea dynamics and ecosystem functioning.
For example, the turbulent flux coefficient is often assumed constant, $\Gamma = 0.2$ (Section \ref{sec:seamix} and \ref{sec:inframix}), but it is unknown if this holds in the wake of OWTs where fine scale density structures and strong spatial variability are present.
Further, production of TKE due to infrastructure has been assumed constant in time and evenly distributed over the area `occupied' by the monopile.
In reality production is focused in the narrow wake of individual monopiles, and could vary by two orders of magnitude during the spring-neap tidal cycle (Section \ref{sec:inframix}). 
TKE production, and thus mixing, arising from stratified flow past infrastructure is also dependent on the drag coefficient $c_D$,
yet little is known about the dependence of $c_D$, particularly at high Reynolds numbers or where mixing length scales are large in comparison to density length scales. 
Meanwhile, research on drag past other obstacle forms suggest estimates for drag may be incorrect by orders of magnitude.
Advancing our understanding of each of these processes is vital for assessing their impact on shelf sea oceanography and ecosystem functioning.
\newline
\subsection{Shelf Sea Dynamics}
\begin{figure}
    \centering
    \includegraphics[width=\textwidth]{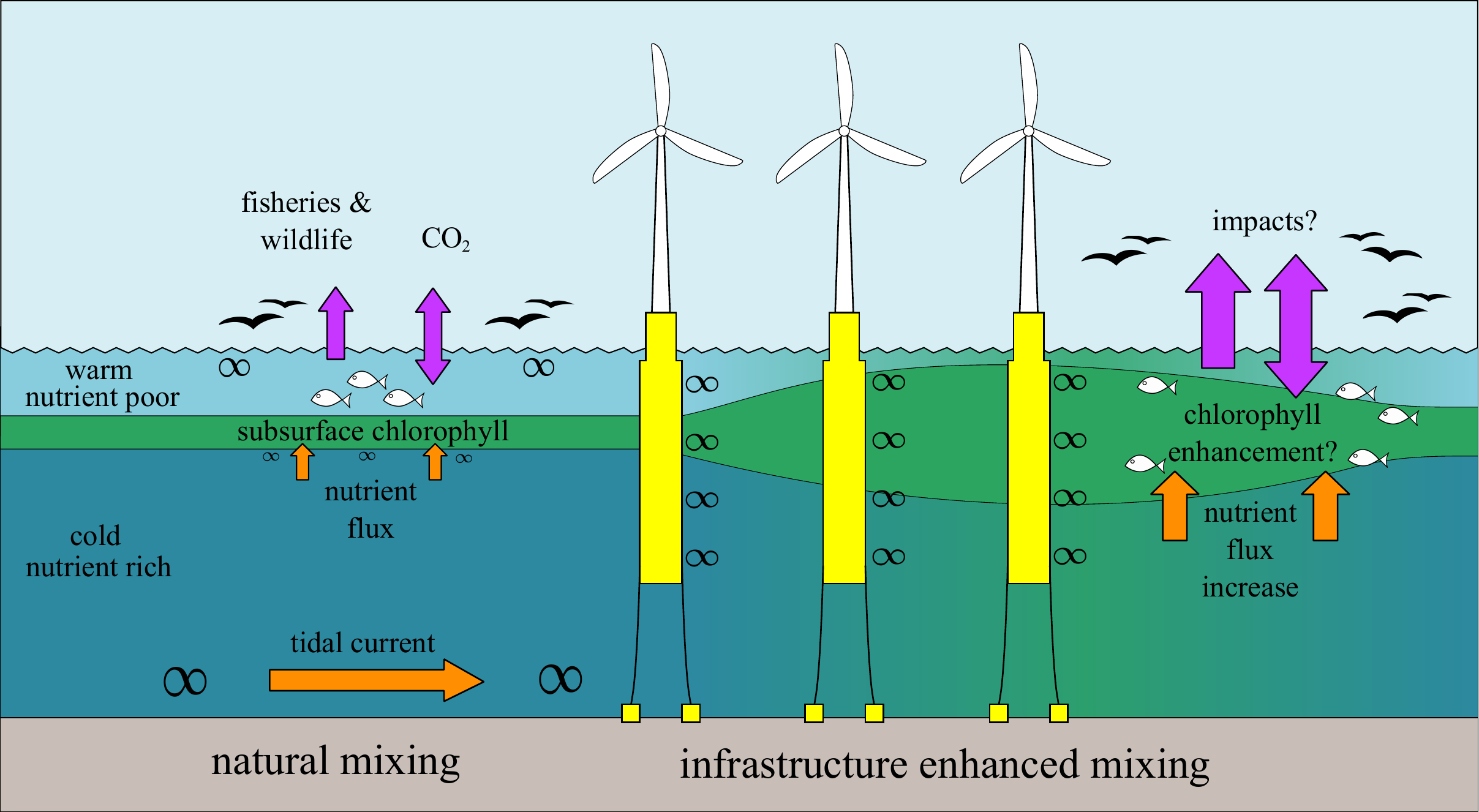}
    \caption{ 
    Offshore wind infrastructure adds wake turbulence throughout the upper water column, including directly at the thermocline.
    Here swirl size indicates turbulence intensity arising from near bed, near surface and flow-structure induced shear.
    Wake turbulence mixes cold nutrient rich bottom water with warm nutrient poor surface water, reducing the strength of stratification and potentially enhancing plankton growth in the subsurface chlorophyll layer.
    Changes in the subsurface chlorophyll layer would have further impacts on nutrient pathways, ecosystem functioning and oceanic carbon sequestration.
    }
    \label{fig:final}
\end{figure}

Offshore wind farms are anticipated in having large local impacts on shelf sea dynamics, in a similar fashion to natural topographically controlled mixing, e.g. driven by internal waves and flow over seafloor sand banks.
It is anticipated that broader and more diffuse thermocline would develop as a result of enhanced mixing, weakening it as a barrier to vertical mixing and transport (Figure \ref{fig:final}). Subsequent change to surface water characteristics would likely alter exchange across the ocean-atmosphere interface, with impacts on heat storage,  atmospheric CO$_2$ uptake and benthic resupply of O$_2$.
The scale of this response will be site and infrastructure specific.
At regional scales the water column should re-stratify subject to natural buoyancy forcing. 
In a more extreme scenario, strongly enhanced mixing could prevent stratification from forming around wind farms.  This effect is observed around islands, where reduced surface temperatures and well-mixed waters result from enhanced mixing due to flow acceleration and seabed shoaling \citep{simpson1982mixing}.  Small, but significant residual currents sweep this well-mixed water into an observable downstream wake.  
\par
Enhanced mixing from infrastructure may also impact seasonal, and shorter timescale, cycles.
The first order response of the vertical density structure to enhanced mixing in a wind farm region would likely be delayed onset and early breakdown of seasonal stratification with weaker stratification throughout the summer season.
Development of near surface stratification during periods of low wind stress would no longer be expected to occur as the enhanced mixing would act to persistently stir the normally episodically mixed surface layer.  
\par
With infrastructure development from local scale of a single turbine to regional scales of multiple developments offshore wind farms in shelf seas have the potential to have regional impact.
Horizontal variation in density may arise from wind farm scale mixing, and these variations may enhance submesoscale processes driving additional vertical transport and mixing.
The density of offshore wind farms, the regional distribution of mixing and wake-wake interactions between wind farms will be of critical importance in determining shelf sea response to offshore wind development.
\newline
\subsection{Shelf-Sea Ecosystem Functioning}
Enhanced mixing rates due to infrastructure, would not only lead to temporal and spatial variations in vertical density structure, but also impact biogeochemical function at a fundamental level that would cascade up the ecosystem.
\par
Stronger mixing in the thermocline will drive more nutrients from the bottom water up into the subsurface chlorophyll maximum and, if the mixing is strong enough, up into the surface layer where it could support additional phytoplankton growth.
The rate of turbulent mixing strongly affects the simulation of ecosystem behaviour, as demonstrated by \citep{Luneva19}, who found that different mixing schemes caused a shift in the spring bloom by 1 month, and change regional chlorophyll differences by order 100\% .
\par
Mixing also alters the light experienced by the phytoplankton, with stronger mixing potentially disrupting light sufficiently to hinder photosynthesis. 
The net effect, i.e. whether or not the extra mixing aids net phytoplankton production, will depend on some balance between the nutrient and light effects. 
The summer reduction in bottom water oxygen concentrations will also respond to the increased mixing. Oxygen will be supplied from the surface water downward, potentially offsetting some of the normal bacterial demand for oxygen as they recycle the sinking organic detritus from phytoplankton growth. 
At the same time, however, there may be an increased supply of organic detrital material if the net effect of mixing on primary production was positive, in which case the bacterial demand for oxygen in the bottom water will increase. 
Understanding the net effect of mixing on bottom water is important, as there are large areas of stratified shelf seas that are currently viewed as being close to experiencing oxygen depletion in late summer \citep{ciavatta2016decadal}.
\par
Changes to biogeochemical functioning would need to be assessed over several years. 
For instance, an immediate positive impact on net phytoplankton production because of the extra nutrient supply will mean that the total amount of the water column inventory of nutrients used for that year will have increased. 
The shelf is not restocked with fresh nutrients from the open ocean every year \citep{ruiz2019seasonality}, so the ability of the shelf system to maintain the increased production will depend on how efficient the system is at recycling organic nutrients back to inorganic nutrients particularly over the winter.  
\par
A natural analogue to the effects of renewable energy infrastructure in a stratified environment may be found in the central Celtic Sea, where a number of seabed banks interact with tidal flows to inject significant internal wave-driven mixing into the thermocline \citep{Palmer2013}. 
While the exact biogeochemical and biological reasons are not yet clear, it does appear that the bank-driven mixing increases the overall biological activity of the region, ultimately resulting in commercial fishing \citep[see, e.g., Figures \ref{fig:globalwind}c and\ref{fig:SCMprofile} ][]{sharples2013fishing}. The island mixing effect too \citep{simpson1982mixing}, produces significantly enhanced nutrient fluxes, with a corresponding increase in plankton production and fishing activity.
Thus, combined with reduced fishing pressure, appropriate infrastructure in suitable locations could prove a positive impact on regional ecology, with benefits to wildlife and fisheries.  
\par
The changes in the timing of stratification expected from increases in mixing also need to be considered. 
Late development of stratification and the spring bloom has the potential to  impact wildlife, which has evolved to take advantage of this abundance, with seabird breeding populations and fish stocks a notable risk. 
At the other end of the stratified period, additional mixing will lead to earlier re-mixing of the whole water column, shortening the total productive time of the area and also altering the timing of the autumn bloom.
Similarly, some seabird colonies are located to take advantage of the enhanced productivity at tidal mixing fronts \citep{Trevail2019a}, the location of which may be affected by additional mixing from offshore wind farm infrastructure.  
\newline
\subsection{Management and Mitigation Requirements}
Environmental Impact Assessments (EIA) are required for all new offshore wind farms to mitigate negative impacts resulting from construction, operation and decommissioning across the lifecycle of use.  
Surveying also covers geophysical site assessment, focus on seabed composition, current assessments and the potential for seabed scour around anchors and cables. 
Current EIAs have been developed for mixed coastal environments to ensure sustainable growth of the offshore wind sector. 
Beyond the construction phase, impact has been focused on the benthic habitat and individual species interactions with infrastructure.  
Impact on marine mammals, seabird colonies, and fisheries in the region are individually assessed, but without consideration of potential alteration of the primary production through enhanced mixing.
\par 
To ensure the continued growth of the sector \citep{broadhorizons}, impacts of the new generation of developments in deeper seasonally stratified regimes will likely require a more fundamental assessment. 
Baseline surveys must include the natural cycle of water column stratification, biogeochemical fluxes, and primary production.  
Accurately forecasting the interaction between the flow, infrastructure and stratification will require site, array and design specific observations and model scenarios.  
Only with a comprehensive understanding of this physical modification to the biogeochemical functioning of shelf seas, can impact throughout the marine web be adequately assessed. 
\section{Conclusions}
Previous work has considered the environmental impact that offshore wind energy has on well-mixed shallow-water marine ecosystems, including from benthic habits, fisheries to seabirds. Much of this work remains relevant to enable sector growth. However, sector expansion from well-mixed shallow water to seasonally stratified deeper water represents a fundamental change, where physical and environmental impacts are not understood.
\par
For the first time planned developments of both fixed and floating offshore wind infrastructure will add large scale anthropogenic mixing to seasonally stratified shelf seas.
Large scale mixing may force shelf sea physics, establishing a `new normal' for biogeochemical cycles and shelf sea ecosystem functioning. 
The potential benefits and risks posed by infrastructure mixing of stratified shelf seas, on top of climate change, represents a combined hazard that has not been considered.
\par
Locally, flow past offshore wind energy infrastructure results in (barotropic) drag and turbulence that, by itself, dissipates at least 140\% more energy than exists naturally at the thermocline (section \ref{sec:inframix}).
However, in stratified waters, the additional baroclinic (wave) drag can exceed the barotropic drag.
For example, the baroclinic drag for a stratified flow past topography can be 1-2 magnitudes larger than the barotropic drag \citep{smith1978measurement}.
Baroclinic drag from vertical infrastructure in stratified flow, such as OWT, is as yet unquantified.
Moreover, the role of horizontal shear on vertical mixing, here produced by obstacles that scale with density gradient length scales, is poorly understood, particularly in complex stratifications like the thermocline.
It is of great importance to extend our understanding of the fundamental fluid dynamics of flows past vertical structures in ocean-realistic stratifications, in particular, the onset of turbulence and ensuing mixing associated with the breakdown of induced vortices.
\par
Regionally, the first order paradigm for seasonally stratified shelf seas is the balance between the stratifying influence of surface heating, and the input of mechanical energy to mix the water column at the upper and lower boundaries (due to wind stress and the tidal shear respectively). 
High resolution shelf sea models have some success in reproducing this \citep{Luneva19}.
However, the addition of mixing from large scale offshore wind farm development, limits our ability to understand the trajectory of shelf sea ecosystems. 
To address this, research is urgently needed that scales processes from: a single turbine; an array of turbines composing a wind farm; to an entire shelf sea region with multiple farms. 
Advances in regional ecosystem modelling must then be validated against direct before-and-after observations to skillfully assess the direct and indirect impacts of anthropogenic mixing, and so guide sustainable development.
\par
Growth of the offshore wind energy industry must be accelerated to meet global 2050 Net Zero commitments. 
Risks posed by offshore wind development in stratified shelf seas should be mitigated against, but potential benefits should be identified and maximised.
Research should consider how nature based solutions, foundation design and array layout can enable sector growth, mitigation of risks and maximisation of benefits. 
\newline
\section*{Acknowledgements}
RMD acknowledges the support of the UK Natural Environment Research Council NE/S014535/1; CJL acknowledges the support of the Offshore Renewable Energy Catapult; BJL acknowledges the support of the Smart Efficient Energy Centre, Bangor University, part funded by the European Regional Development Fund; DMG acknowledges the support of the UK Engineering and Physical Sciences Research Council EP/S000747/1; and JAP acknowledges the support of the Natural Environment Research Council Climate Linked Atlantic Sector Science (CLASS) programme.
\newline
\bibliographystyle{frontiersinSCNS_ENG_HUMS}
\bibliography{references.bbl}

\end{document}